\documentclass[aps,twocolumn,showpacs,byrevtex]{revtex4-1}
\usepackage[colorlinks,linkcolor=blue,anchorcolor=blue,urlcolor=blue,citecolor=blue]{hyperref}
\usepackage{lineno}
\usepackage{times}
\usepackage{multirow}
\usepackage{amsmath}
\usepackage{graphicx}
\usepackage{float}
\usepackage{color}

\usepackage[T1]{fontenc}

\newcommand{\kk}{K^+K^-}

\newcommand{\ee}{e^+e^-}

\newcommand{\jpsi}{J/\psi}
\newcommand{\piz}{\pi^0}
\newcommand{\chicz}{\chi_{c0}}

\newcommand{\ar}{\rightarrow}
\newcommand{\ggphi}{\gamma\gamma\phi}

\begin{document}

\title{\boldmath Partial-wave analysis of $\jpsi\ar\gamma\gamma\phi$
}

    \author{
    M.~Ablikim$^{1}$, M.~N.~Achasov$^{5,b}$, P.~Adlarson$^{74}$, X.~C.~Ai$^{80}$, R.~Aliberti$^{35}$, A.~Amoroso$^{73A,73C}$, M.~R.~An$^{39}$, Q.~An$^{70,57}$, Y.~Bai$^{56}$, O.~Bakina$^{36}$
% ...
    	\\
    	\vspace{0.2cm}
    	(BESIII Collaboration)\\
    	\vspace{0.2cm} {\it
    		$^{1}$ Institute of High Energy Physics, Beijing 100049, People's Republic of China\\
% ...
$^{5}$ Budker Institute of Nuclear Physics SB RAS (BINP), Novosibirsk 630090, Russia\\
% ...
$^{35}$ Johannes Gutenberg University of Mainz, Johann-Joachim-Becher-Weg 45, D-55099 Mainz, Germany\\
$^{36}$ Joint Institute for Nuclear Research, 141980 Dubna, Moscow region, Russia\\
% ...
$^{39}$ Liaoning Normal University, Dalian 116029, People's Republic of China\\
% ...
$^{56}$ Southeast University, Nanjing 211100, People's Republic of China\\
$^{57}$ State Key Laboratory of Particle Detection and Electronics, Beijing 100049, Hefei 230026, People's Republic of China\\
% ...
$^{70}$ University of Science and Technology of China, Hefei 230026, People's Republic of China\\
% ...
$^{73}$ University of Turin and INFN, (A)University of Turin, I-10125, Turin, Italy; (B)University of Eastern Piedmont, I-15121, Alessandria, Italy; (C)INFN, I-10125, Turin, Italy\\
$^{74}$ Uppsala University, Box 516, SE-75120 Uppsala, Sweden\\
% ...
$^{80}$ Zhengzhou University, Zhengzhou 450001, People's Republic of China\\
\vspace{0.2cm}
% ...
$^{b}$ Also at the Novosibirsk State University, Novosibirsk, 630090, Russia\\
% ...
    	}
        }

\date{\today}

\begin{abstract}

Using a sample of $(10087\pm44)\times10^{6}$ $\jpsi$ events collected by the BESIII detector at the BEPCII collider, a partial-wave analysis of the decay $\ggphi$ is performed to investigate the intermediate resonances in $\jpsi\ar\gamma R, R\ar\gamma\phi$. Resonances including $f_{1}(1285)$, $\eta(1405)$, $f_{1}(1420)$, $f_{1}(1510)$, $f_{2}(1525)$, $X(1835)$, $f_{2}(1950)$, $f_{2}(2010)$, $f_{0}(2200)$ and $\eta_{c}$ have been detected with statistical significance greater than 5$\sigma$. The product branching fractions $\mathcal{B}(\jpsi\ar\gamma R, R\ar \gamma \phi)$ are reported, and the resonance parameters of $\eta(1405)$ and $X(1835)$ have been precisely determined. It is more probable that $\eta(1405)$ and $X(1835)$ act as the first - and second - excited states of $\eta^{\prime}$, respectively. The observation of $\eta_{c}\ar\gamma\phi$ holds great significance for the in depth study of charmonium properties. No evidence of $\eta(1295)$, $\eta_{1}(1855)$ or X(2370) is found. The measured upper limits of relevant BF products do not conflict with theoretical expectations for X(2370) as a pseudoscalar glueball and $\eta_{1}(1855)$ as a hybrid state.
\end{abstract}


\maketitle
\section{Introduction}
The non-abelian nature of quantum chromodynamics (QCD) predicts the existence of various exotic states, including glueballs, hybrids, and multiquark states~\cite{intro0}. Confirming these states experimentally would provide crucial insights into the confinement regime of QCD and serve as a direct test of QCD theory. Radiative decays of charmonium are glue-rich processes and serve as excellent probes for investigating the production of gluonic matter and light hadron structures~\cite{intro1}. While significant progress has been made in the past few decades~\cite{intro2}, numerous unresolved issues persist.

One of the current focal points is the nature of the two pseudoscalar states around 1.4 GeV/$c^2$~\cite{intro0}, namely, the $\eta(1405)$ and $\eta(1475)$ as listed by the Particle Data Group (PDG)~\cite{ref1}. 
The $\eta(1475)$ is considered a potential first radial excitation of the $\eta^{\prime}$, while the $\eta(1405)$ has been suggested as a candidate for the ground state pseudoscalar glueball ~\cite{discuss,puzzle}. 
However, the mass of $\eta(1405)$ is relatively far from the predicted mass of a pseudoscalar glueball (2.3\textendash2.6 GeV/$c^2$) by Lattice QCD~\cite{ref2,ref3}, leading to the long standing “E-$\iota$ puzzle”~\cite{eiota}. It is still controversial whether the $\eta(1405)$ and $\eta(1475)$ are two separate states or just one pseudoscalar state observed in different decay modes.

The $X(1835)$ was first observed in the decay $\jpsi \ar \gamma\pi^{+}\pi^{-}\eta^{\prime}$ by BESII~\cite{ref4}, with its spin-parity quantum numbers established as $J^{PC} = 0^{-+}$ by BESIII~\cite{ref5}. This discovery led to various theoretical speculations regarding its nature, including  proposals such as $N\bar{N}$ bound state~\cite{ref6}, a baryonium with a sizable gluon content~\cite{ref7}, a second radial excitation of the $\eta^{\prime}$~\cite{ref8,ref9}, an $\eta_{c}$-glueball hybrid, and a pseudoscalar glueball~\cite{ref10}. So far, none of these interpretations has been ruled out or confirmed. 

The $X(2370)$ has also been observed in the $\jpsi \ar \gamma\pi^{+}\pi^{-}\eta^{\prime}$~\cite{ref101,ref11}. Recently, the BESIII experiment confirmed that the spin-parity of the $X(2370)$ as $0^{-+}$ via the decay $\jpsi\ar\gamma K_{S}^{0}K_{S}^{0}\eta$~\cite{ref2370}. This experimental finding has prompted numerous theoretical speculations regarding its nature~\cite{ref8,2370p2,2370p3,2370p4,2370p5},  with one fascinating hypothesis being a pseudoscalar glueball~\cite{ref3,refLQCD2,refLQCD3,refLQCD4,refLQCD5}. By using the radiative pseudoscalar glueball decay width, which is approximately $\Gamma_{PS\rightarrow\gamma\phi}\sim$95 keV, the total pseudoscalar
glueball decay width is approximately $\Gamma_{PS}\sim 470...614 $ MeV~\cite{2370}, and the braching fraction (BF) of $\jpsi$ radiatively decaying into pseudoscalar glueball being $2.31(80)\times10^{-4}$~\cite{puzzle}, it can be deduced that if the $ X(2370)$ is a pseudoscalar glueball, the $ \mathcal{B}(\jpsi\ar\gamma X(2370)\ar\gamma\gamma\phi)$ should be less than $10^{-8}$.

Recently, the BESIII collaboration reported the first observation of a $I^{G}(J^{PC})=0^{+}(1^{-+})$ structure $\eta_{1}(1855)$ through a partial-wave analysis (PWA) of the $\jpsi\ar\gamma\eta\eta^{\prime}$ decay~\cite{1855}, sparking a variety of theoretical interpretations, including hybrids~\cite{hybrid1,hybrid2,hybrid3}, molecular states~\cite{molecule1,molecule2}, and tetraquark states~\cite{tetraquark1,tetraquark2}. When considering the $\eta_{1}(1855)$ as an $s\bar{s}g$ hybrid, the total decay width is predicted to be approximately 160 MeV~\cite{1855_0}, and the radiative decay width $\Gamma_{\eta_{1}(1855)\ar\gamma\phi}\sim4.2\times 10^{-2}$~\cite{1855_2}. Given that the BF for the $\jpsi\ar\gamma\eta_{1}(1855)$ decay is on the order of $10^{-5}$~\cite{1855_3}, the BF $\mathcal{B}(\jpsi\ar\gamma\eta_{1}(1855),\eta_{1}(1855)\ar\gamma\phi)$ should be less than $10^{-9}$.

Since radiative decays do not change the flavor structure, the decays $\jpsi\ar\gamma R$, $R\ar\gamma V$ with $V \equiv \rho, \omega, \phi$, serve as flavor filter reactions and play an important role in unraveling the quark contents of the intermediate resonances~\cite{ref12,ref13}. Throughout the text, the  symbol $R$ denotes intermediate resonance. Previously, the BESIII collaboration has studied the prominent features of $f_{1}(1285)$, $\eta(1475)$, and $X(1835)$ using 1.3 billion $\jpsi$ events~\cite{refkang}. A one-dimensional fit to the $\gamma\phi$ invariant mass ($M(\gamma\phi)$) distribution from $\jpsi\ar\gamma\gamma\phi$ was performed.
In addition, from the angular distribution it was found that the data favors spin and parity ($J^{PC}$) assignments of $0^{-+}$ for the structures around 1.4 GeV/$c^{2}$ and 1.8 GeV/$c^{2}$.

Using a dataset of $(10087\pm44)\times10^{6}$ $\jpsi$~\cite{Njpsi} events from $\ee$ annihilation data collected at the center-of-mass (c.m.) energy $\sqrt{s}=3.097$ GeV with the BESIII detector at the BEPCII collider, we present the results of a PWA of $\jpsi\ar\gamma\gamma\phi$ to extract the contributions of intermediate components and determining their spin-parity, mass, and width. 
The research results will provide useful information to help us understand the properties of $\eta(1405/1475)$ and $X(1835)$.

\section{BESIII Detector and MC Sample}
The BESIII detector~\cite{Ablikim:2009aa} records symmetric $e^+e^-$ collisions from the BEPCII storage ring~\cite{Yu:IPAC2016-TUYA01} in the c.m. energy range from 1.84 to 4.95~GeV, achieving a peak luminosity of $1.1 \times 10^{33}\;\text{cm}^{-2}\text{s}^{-1}$ at $\sqrt{s} = 3.773\;\text{GeV}$. BESIII has collected large data samples in this energy region~\cite{puzzle,EcmsMea,EventFilter}. The cylindrical core of the BESIII detector covers 93\% of the full solid angle and includes a helium-based multilayer drift chamber~(MDC), a plastic scintillator time-of-flight system~(TOF), and a CsI(Tl) electromagnetic calorimeter~(EMC), which are all enclosed in a superconducting solenoidal magnet providing a 1.0~T magnetic field. The magnetic field was 0.9~T in 2012, which affects 10.8\% of the total $J/\psi$ data. The end cap TOF system was upgraded in 2015, affecting 87.0\% of the data used in this analysis ~\cite{etof}. A detailed description of the BESIII detector design and performance can be found in Ref.~\cite{Ablikim:2009aa}.
~\

Simulated data samples, generated with a {\sc geant4}-based~\cite{geant4} Monte Carlo (MC) package, that includes the geometric description of the BESIII detector and the detector response~\cite{BESIIIDetector}, are utilized to determine detection efficiencies and estimate backgrounds. The simulation includes modeling of the beam energy spread and initial state radiation in the $e^+e^-$ annihilations using the generator {\sc kkmc}~\cite{ref:kkmc}. Signal MC samples of the process $\ggphi$ with the subsequent decays $\phi\ar\kk$ are generated uniformly in phase space (PHSP). The inclusive MC sample includes both the production of the $J/\psi$ resonance and the continuum processes incorporated in {\sc kkmc}. All particle decays are modeled with {\sc evtgen}~\cite{ref:evtgen} using BFs either taken from the PDG~\cite{ref1}, when available, or otherwise estimated with {\sc lundcharm}~\cite{ref:lundcharm}. Final state radiation from charged final state particles is incorporated using the {\sc photos} package~\cite{photos}.

\section{Event Selection}\label{sec:selection}
Charged tracks detected in the MDC are required to be within a polar angle ($\theta$) range of $|\rm{cos\theta}|<0.93$, where $\theta$ is defined with respect to the $z$-axis, the symmetry axis of the MDC. The distance of closest approach to the interaction point (IP) must be less than 10\,cm along the $z$-axis and less than 1\,cm in the transverse plane. Particle identification for charged tracks combines measurements of the d$E$/d$x$ in the MDC and the flight time in the TOF to form likelihoods $\mathcal{L}(h)~(h=p,K,\pi)$ for each hadron $h$ hypothesis. A track is identified as a kaon if its kaon hypothesis likelihood satisfies $\mathcal{L}(K)> \mathcal{L}(\pi)$ and $\mathcal{L}(K)> \mathcal{L}(p)$.

Photon candidates are identified by the showers produced in the EMC. The deposited energy of each shower must be larger than 25~MeV in the barrel region ($|\cos \theta|< 0.80$) and larger than 50~MeV in the end cap regions ($0.86 <|\cos \theta|< 0.92$). To exclude showers that originate from charged tracks, the angle subtended by the position of EMC shower~\cite{emcshower} and the position of the closest charged track at the EMC as measured from the IP must be greater than 10 degrees as measured from the IP. To suppress electronic noise and showers unrelated to the event, the difference between the EMC time and the event start time is required to be within [0, 700]\,ns. Candidate events are required to have two oppositely charged tracks identified as kaons and have at least two photons.

A four-constraint (4C) kinematic fit is performed under the~$\jpsi\ar\gamma\gamma K^{+} K^{-}$ hypothesis for all combinations with good photon candidates. For events with more than one combination of $\jpsi\ar\gamma\gamma K^{+} K^{-}$, only the combination with the minimum $\chi^{2}_{4\rm C}$ is retained. To suppress potential multi-photon backgrounds, additional 4C kinematic fits for $\gamma\kk$, $3\gamma\kk$, or $4\gamma\kk$ hypotheses are performed. An event is retained for further analysis only if $\chi^{2}_{4\rm C}(\gamma\gamma\kk)$ is smaller than $\chi^{2}_{4\rm C}$ of any additional hypotheses. Moreover, the $\chi^{2}_{4\rm C}$  resulting from the 4C kinematic fit must be less than 40.

\begin{figure}[!htp]
 \begin{center}
     \includegraphics[width=0.23\textwidth]{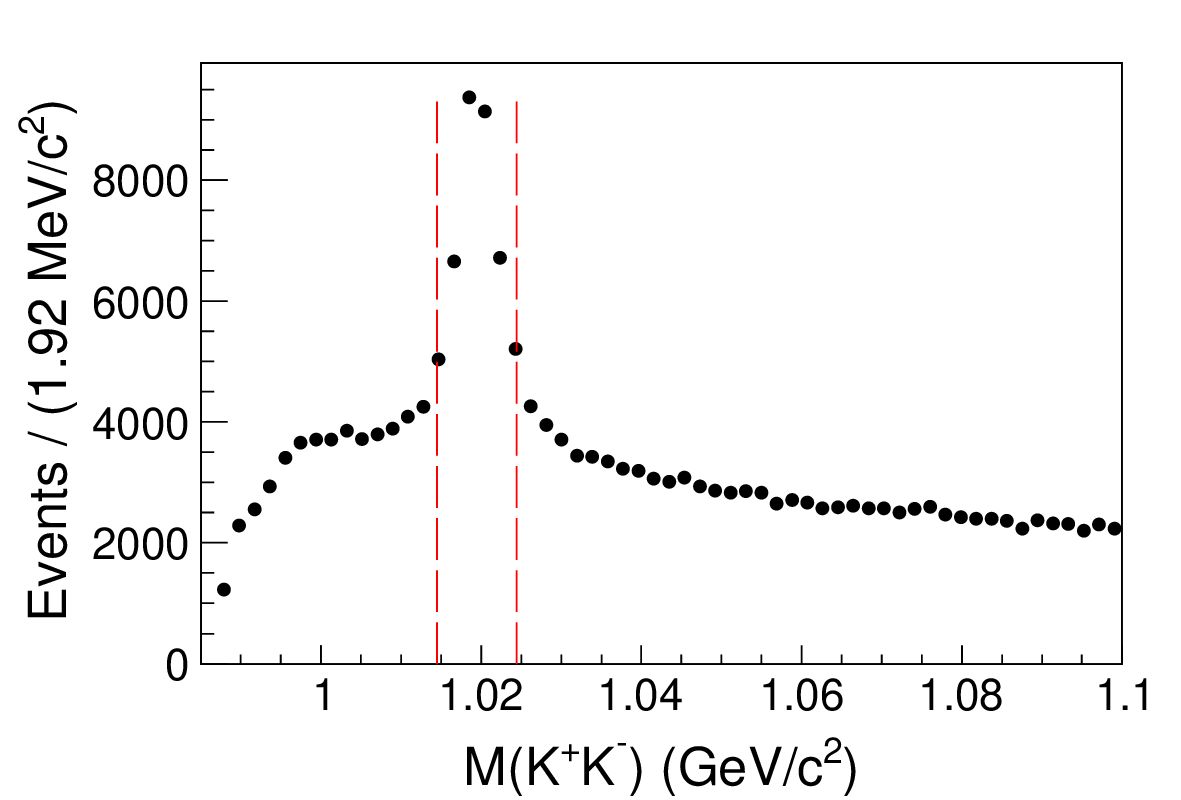}
     \put(-30,60){(a)}
     \includegraphics[width=0.24\textwidth]{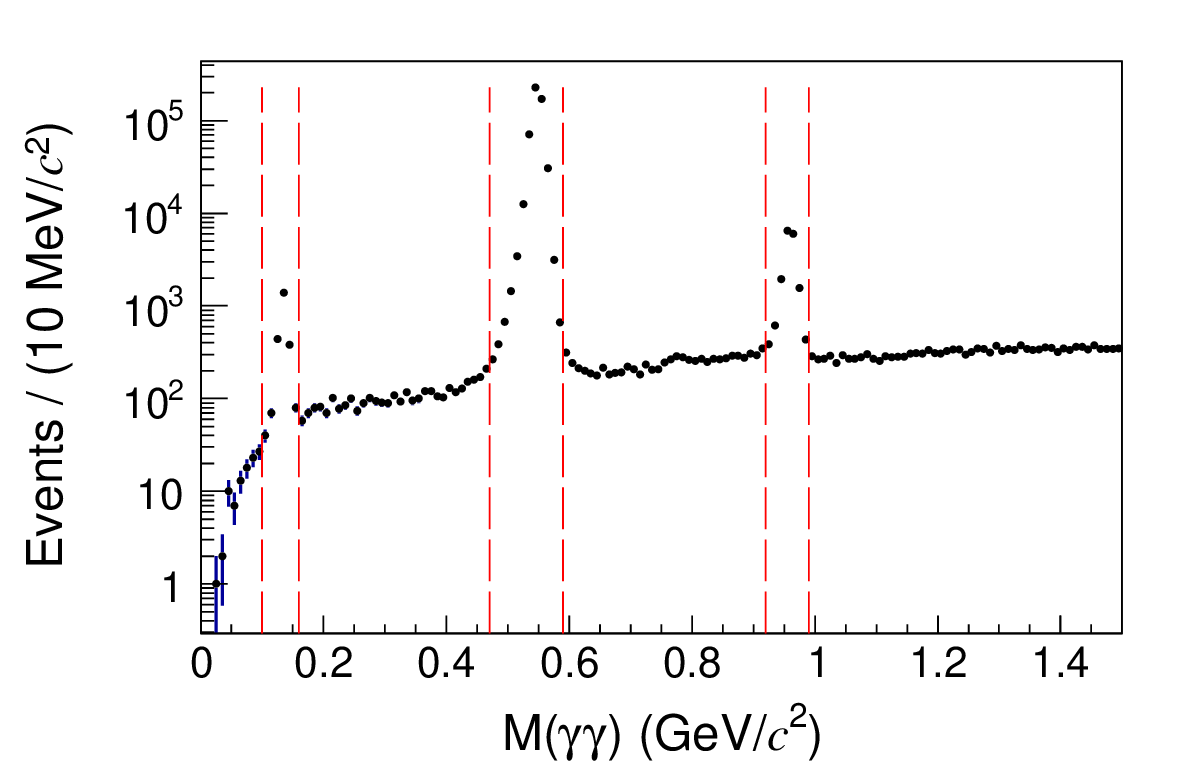}
     \put(-30,60){(b)}
     \caption{Distributions of (a) $M(\kk)$ and (b) $M(\gamma\gamma)$ with $M(\kk)$ in the $\phi$ mass window of the accepted $\jpsi\ar\gamma\gamma\phi$ candidate events in data, where the pair of red dashed vertical lines in the left figure denote the $\phi$ signal region, while the pairs of red dashed vertical lines in the right figure denote the $\pi^0$, $\eta$ or $\eta^{\prime}$ background regions}
    \label{kkdistribution}
 \end{center}
 \end{figure}

After the initial event selection, the $M(K^{+}K^{-})$ distribution is shown in Fig.~\ref{kkdistribution}(a). The $\phi$ candidates are identified by requiring $|M(\kk)-M_{\phi}|<0.005$~GeV$/c^2$, where $M_{\phi}$ is $\phi$ nominal mass. Figure~\ref{kkdistribution}(b) shows the $M(\gamma\gamma)$ 
distribution in the $\phi$ signal region, in which clear peaks corresponding to $\pi^0$, $\eta$, and $\eta^{\prime}$ can be seen.
To eliminate backgrounds involving $\pi^{0}\ar\gamma\gamma$, $\eta\ar\gamma\gamma$, or $\eta^{\prime}\ar\gamma\gamma$ decays, events with $M(\gamma\gamma)$ within the ranges $M(\gamma\gamma)\in(0.11, 0.16)$~GeV/$c^{2}$, $M(\gamma\gamma)\in(0.46, 0.59)$ GeV/$c^{2}$ or $M(\gamma\gamma)\in(0.92, 0.99)$ GeV/$c^{2}$ are excluded.

\begin{figure}[!htp]
\center
\includegraphics[width=0.24\textwidth]{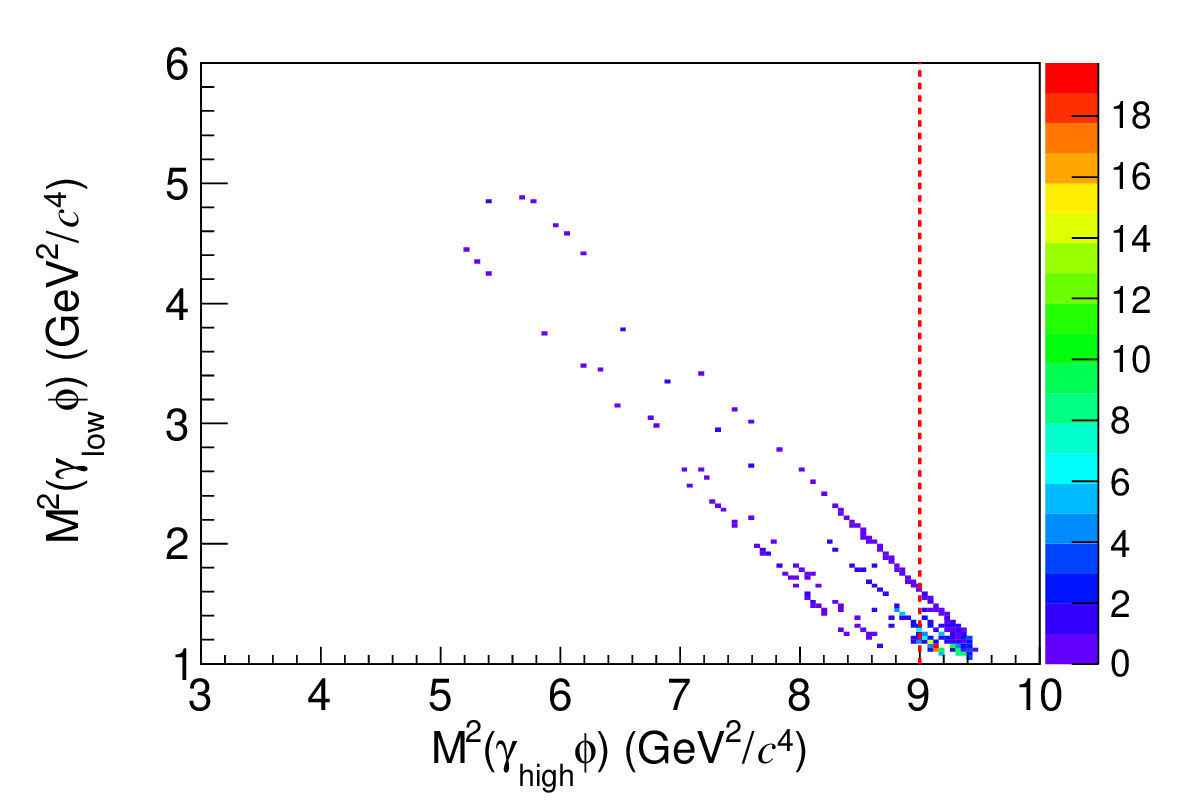}
\put(-40,60){(a)}
\includegraphics[width=0.24\textwidth]{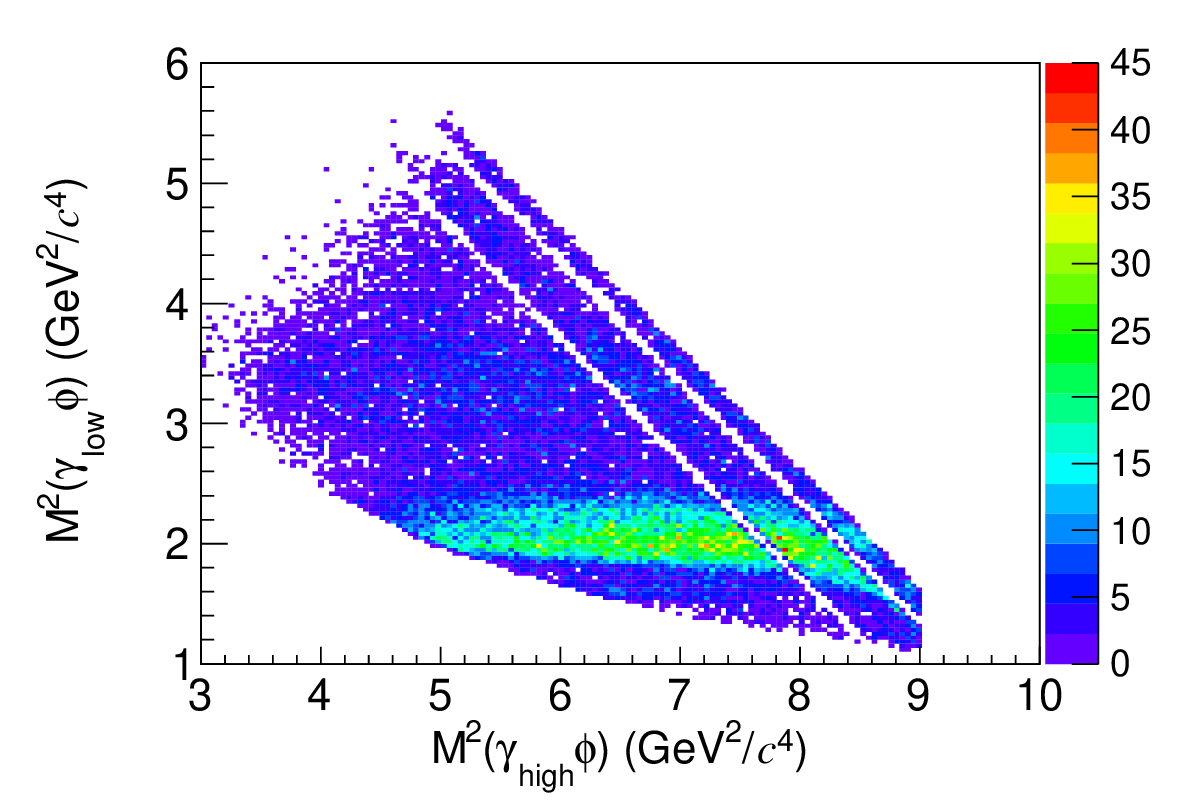}
\put(-40,60){(b)}
\caption{(a) Dalitz plot of $M^{2}(\gamma_{\text high}\phi)$ versus $M^{2}(\gamma_{\text low}\phi)$ for the residual background events of $\jpsi\ar\phi\pi^{0}$, $\jpsi\ar\phi\eta$, $\jpsi\ar\phi\eta^{\prime}$ using the inclusive MC. (b) Dalitz plot of $M^{2}(\gamma_{\text high}\phi)$ versus $M^{2}(\gamma_{\text low}\phi)$ of data following all selection criteria. }
\label{ggremain}
\end{figure}

 Despite these mass window constraints, the backgrounds from $\jpsi\ar\phi\pi^{0}$, $\jpsi\ar\phi\eta$, $\jpsi\ar\phi\eta^{\prime}$ cannot be entirely eliminated due to their tails. We try to quantify residual backgrounds using an inclusive MC sample. Figure~\ref{ggremain}(a), from this sample, depicts the Dalitz plot of $M^{2}(\gamma_{\text high}\phi)$ versus $M^{2}(\gamma_{\text low}\phi)$, where $\gamma_{\text low}$ and $\gamma_{\text high}$ refer to the photons with lower and higher energies in $\jpsi$ rest frame. Notably, Figure~\ref{ggremain}(a) reveals that these events cluster at the high end of the Dalitz plot. By imposing the condition $M^{2}(\gamma_{\text high}\phi)<9$~GeV$^2/c^4$ on  data sample, these backgrounds can be effectively suppressed.
 
After applying all selection criteria to the data, the Dalitz plot of $M^{2}(\gamma_{\text high}\phi)$ versus $M^{2}(\gamma_{\text low}\phi)$ of the accepted $J/\psi\ar \gamma\gamma \phi$ candidate events of data is  depicted in Fig.~\ref{ggremain}(b). Subsequently, 38383 events survive the event selection criteria.

\section{Background Treatment}
\begin{figure}[!htp]
\center
\includegraphics[width=0.35\textwidth]{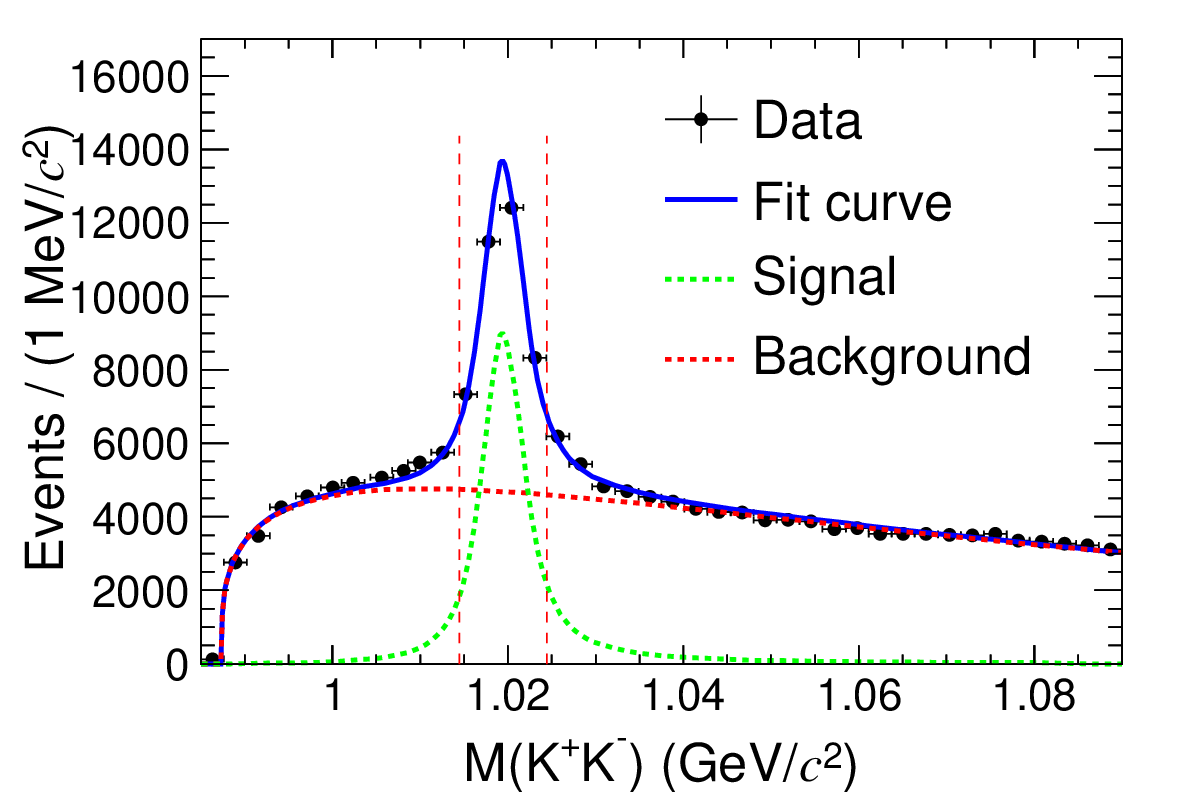}
\caption{The one-dimensional fit to the $M(K^{+}K^{-})$ distribution. The dots with error bars are data, the blue line denotes the fit curve, the green dash line denotes the signal shape, the red dash line is background shape, and the pair of the red dashed vertical lines denote the $\phi$ signal region.}
\label{fitkk}
\end{figure}

To quantify the contributions of $\phi$-events and non-$\phi$-events in the $\phi$ signal region, directly influencing the purity of the $\phi$ sample used in subsequent analyses, a one-dimensional fit to the $M(K^{+}K^{-})$ distribution is performed, as shown in Fig.~\ref{fitkk}. In the fit, the signal is modeled by a relativistic P-wave Breit-Wigner (BW) function convolved with a Gaussian function describing the resolution difference between data and MC simulation. The relativistic P-wave Breit-Wigner function is expressed as
\begin{equation}
 BW(m)=\frac{F_{m}}{m^{2}_{0}-m^{2}-im_{0}\Gamma(m)},
\label{PBW}
\end{equation}
where $F_{m}$ is the Blatt-Weisskopf damping factor $F_{m}=\sqrt{1+(Rq_{0})^{2}}/\sqrt{1+(Rq)^{2}}$, $R$ is fixed to 1.5~GeV$^{-1}$, and
\begin{equation}
 \Gamma(m)=\Gamma(m_{0})(\frac{q}{q_{0}})^{2J+1}(\frac{m_{0}}{m})F^{2}_{m}.
\label{Gamma}
\end{equation}
Here $m_{0}$ is the $\phi$ nominal mass from the PDG, $q$ is momentum of either $K^{+}$ or $K^{-}$ in the $\phi$ rest frame and $q_{0}$ is $q$ evaluated at $m = m_{0}$.

The background shape is parametrized by a low-$x$ cutoff function,

\begin{equation}
 f(x, m_{0},c,p)=x\cdot\left[(\frac{x}{m_{0}})^{2}-1\right]^{p}\cdot \text{exp}\left[ c\cdot\left((\frac{x}{m_{0}})^{2}-1\right)\right].
\label{inargus}
\end{equation}
Here, $x$ represents the observable $M(\kk)$, with $m_{0}$=0.987354 denoting the threshold for two kaons,  while $c$ and $p$ are free parameters in the fit.

Finally, the fractions of $\phi$-events and non-$\phi$-events are determined to be 55.3\% and 44.7\%, respectively. With the event selection criteria mentioned above, the residual backgrounds are classified into two categories: backgrounds without $\phi$ and backgrounds with $\phi$. The handling methods for these two types of backgrounds are as follows.
\subsection{Treatment of non-$\phi$ background}
The dominant non-$\phi$ background components are mainly from $\jpsi\ar\gamma R$ ($R = f_{1}(1285)$, $\eta(1405)$ and $f_{1}(1420)$), $R\ar\kk\pi^{0}$, with the $\gamma K^{+}K^{-}\pi^0$ final state. It is difficult to distinguish the signal from these background components  because the $M(\kk)$ value from $R\ar\kk\pi^{0}$ is close to the $\kk$ mass threshold.

Based on the analysis of the signal-to-background ratio $Q$ in the available PHSP around each specific event, the quality factor method~\cite{qfactor} is used to estimate the non-$\phi$ backgrounds. For this procedure, we need to know the functional dependence of the signal and background distributions in terms of a distinct kinematic variable ($\zeta_{r}$). The first step is to assign a number of $n_{c}$ nearest neighbors for each event including the event itself. In order to measure distances between events, a collection of kinematic observables ($\vec \zeta$) that span the PHSP for the reaction is selected to define a metric. Using this metric, the distance between any two events, $d_{ij}$, is given as
     \begin{equation}
     d_{ij}^{2} = \sum_{k  }\left[\frac{\zeta_{k}^{i}-\zeta_{k}^{j}}{\bigtriangleup_{k}}\right]^{2}~,
     \end{equation}
where the sum is over all kinematic observables ($\vec\zeta$), and $\Delta_{k}$ is a normalization factor that represents the weight of the kinematic variable. 

The $n_{c}$ nearest neighbors are then fit with the unbinned maximum likelihood method, with estimators written as 
\begin{equation}
f(\zeta_{r})=f_{s}\cdot S(\zeta_{r}) +(1-f_{s})\cdot B(\zeta_{r}),
\label{eq2}
\end{equation}
where $S(\zeta_{r})$ and $B(\zeta_{r})$ represent the probability density functions of signal and background, respectively, and $f_{s}$ describes a signal fraction with a value constrained within the range $[0.0, 1.0]$. The Q-value for each event is then given by 
\begin{equation}
Q=\frac{f_{s}\cdot S(\zeta_{r})}{f_{s}\cdot S(\zeta_{r}) + (1-f_{s})\cdot B(\zeta_{r})},
\label{eq3}
\end{equation}
in which Q and (1-Q) represent the probabilities of an event to be signal and background, respectively, which are used as an event-weight to determine the contribution of the signal to a given physical distribution. 

In this analysis, seven coordinates are chosen for the metric, which are presented in Table~\ref{reference}. Throughout this paper, the polar angle is denoted by $\theta$, while the azimuth is represented by $\varphi$. It is necessary to make the value of $n_{c}$ as small as possible. This is to ensure that the phase space (PHSP) cell of all the selected neighbors remains small and that the background varies smoothly within the cell. However, $n_{c}$ also has to be large enough to guarantee stable and reliable fits to $M(\kk)$. Through the study of the Monte Carlo (MC) samples, we have reached the conclusion that setting $n_{c}=200$ will result in an appropriate size.

\begin{table*}[!htp]
\centering
\caption{Set of the kinematic observables ($\vec{\zeta_{k}}$) used for background subtraction.}\label{reference}
\begin{tabular}{lcc}
\hline
\hline
$\vec{\zeta_{k}}$    &Description & $\Delta_{k}$     \\
\hline
$\cos\theta_{\gamma_{\text high}}$ & Cosine of polar angle of the high-energy photon in the rest frame of $\jpsi$&2\\
$\cos\theta_{K^{+}}$   & Cosine of polar angle of $K^{+}$ in the rest frame of $\phi$  &2 \\
$\cos\theta_{\phi}$ & Cosine of polar angle of $\phi$ in the rest frame of $\gamma_{\text low}\phi$ &2 \\
$\varphi_{K^{+}}$ & Azimuth of $K^{+}$ in the rest frame of $\phi$ &2$\pi$\\
$\varphi_{\phi}$& Azimuth of $\phi$ in the rest frame of $\gamma_{\text low}\phi$ &2$\pi$\\
$M^2(\gamma_{\text low}\phi)$ &Invariant mass squared of $\gamma_{\text low}\phi$ &0.1 GeV$^{2}/c^{4}$\\
$M^2(\gamma_{\text high}\phi)$  &Invariant mass squared of $\gamma_{\text high}\phi$ &0.15 GeV$^{2}/c^{4}$ \\
\hline
\hline
\end{tabular}
\end{table*}

\begin{figure}[!htp]
   \centering
   \includegraphics[width=0.46\textwidth]{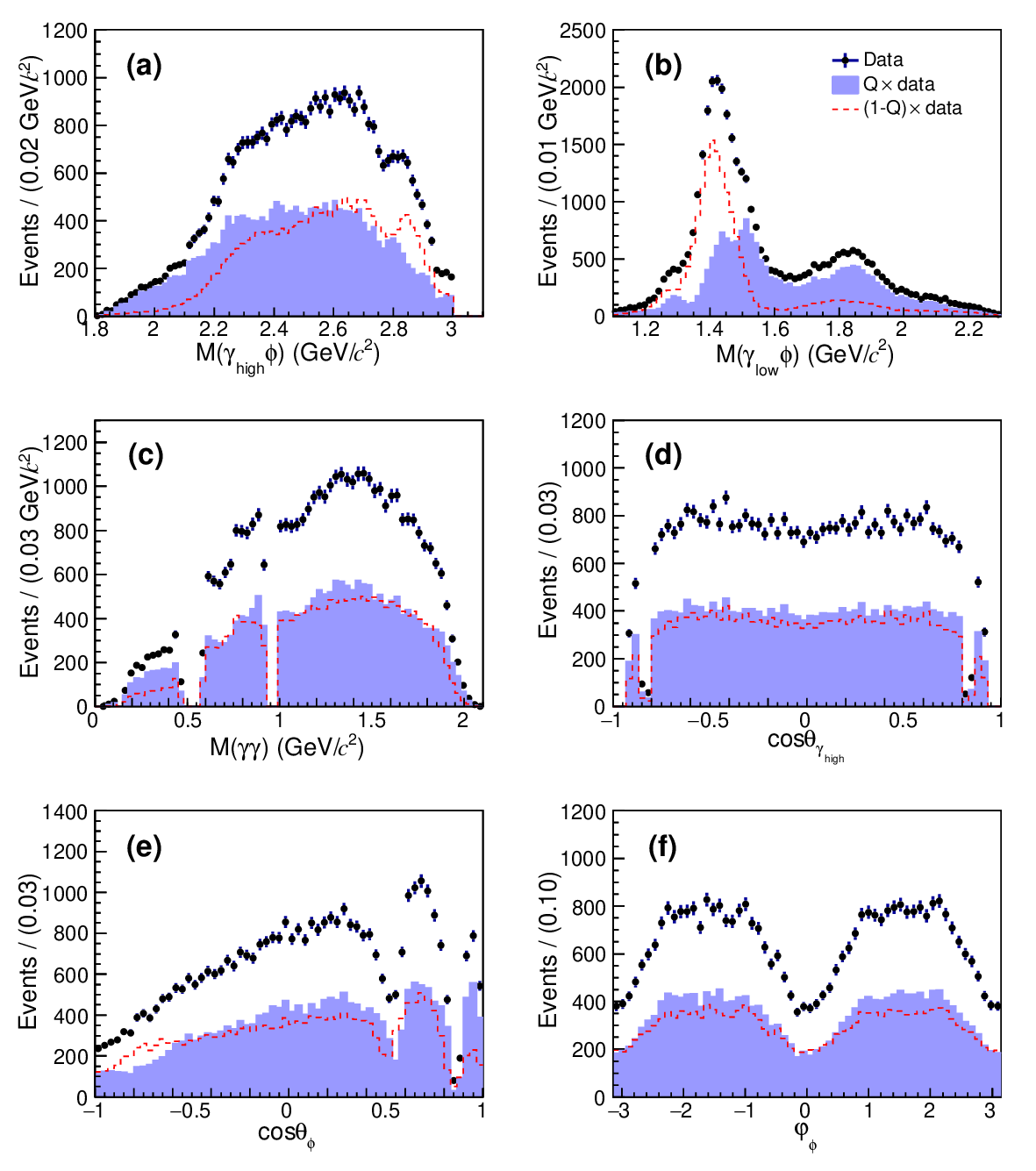}
   \caption{Distributions of (a) $M(\gamma_{\text high}\phi)$, (b) $M(\gamma_{\text low}\phi)$, (c) $M(\gamma\gamma)$, (d) cosine of the polar angle of $\gamma_{\text high}$, (e) cosine of the polar angle of $\phi$ in $\gamma_{\text low}\phi$, and (f) azimuth of $\phi$ of the accepted $J/\psi\ar \gamma\gamma\phi$ candidate events. The black dots with error bars are the selected events in the $\phi$ signal region without any weight, the purple shaded regions are $Q$-weighted data, and the red dashed lines are $(1-Q)$ weighted data.}
   \label{2}
\end{figure}

For this analysis, $M(\kk)$ is chosen as the distinct kinematic variable. The fitting model of $M(\kk)$ is the same as the fit mentioned earlier. As a result, the angular distributions and several invariant mass spectra of the selected events within the $\phi$ signal region with Q-weighted and (1$\mbox{-}$Q)$\mbox{-}$weighted are shown in Fig.~\ref{2}. The sum of all obtained Q-values (weights) is taken as the number of signal events, yielding 20453 events. All further analysis steps are performed using this weighted data sample.

\subsection{Treatment of $\phi$ background}
\begin{figure}[!htp]
   \centering
   \includegraphics[width=0.46\textwidth]{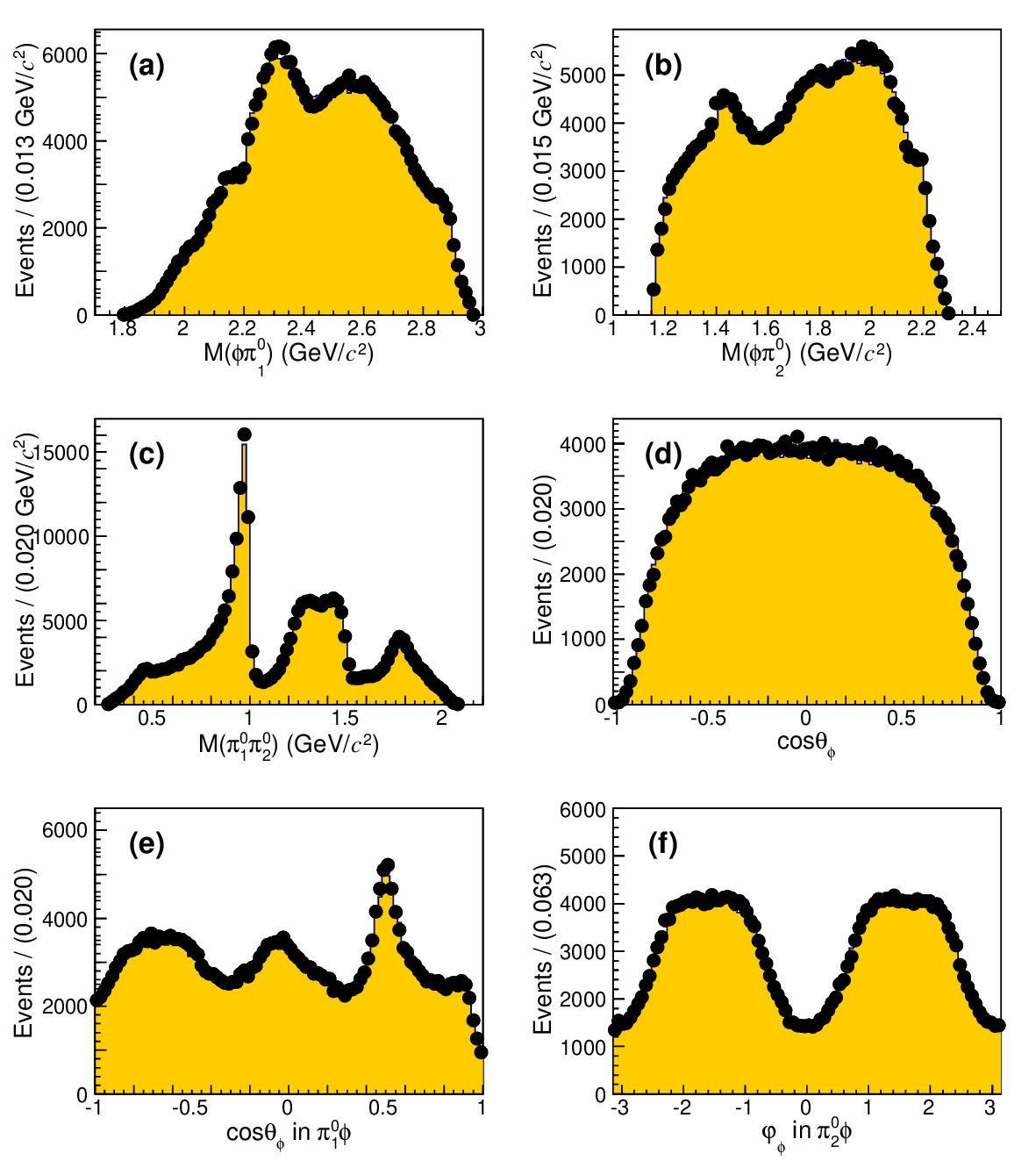}
   \caption{Distributions of (a) $M(\phi\pi^{0}_{1})$, (b) $M(\phi\pi^{0}_{2})$, (c) $M(\pi^{0}_{1}\pi^{0}_{2})$, (d) cosine of the polar angle of $\phi$ in the $\jpsi$ rest frame, (e) cosine of the polar angle of $\phi$ in the $\pi^{0}_{1}\phi$ rest frame, and (f) azimuth of $\phi$ in the $\pi^{0}_{2}\phi$ rest frame of the $\jpsi\ar\phi\pi^{0}\pi^{0}$ candidate events in data (dots with error bars are data) and the re-weighted MC sample (orange shadowed regions).}
   \label{1}
\end{figure}

The dominant $\phi$ related background is $\jpsi\ar\phi\pi^{0}\pi^{0}$, which includes intermediate resonances~\cite{phipipi}. A multi-dimensional re-weighting method~\cite{reweight} is applied to the PHSP MC sample to obtain a ``data-like'' MC sample 
of $\jpsi\ar\phi\pi^{0}\pi^{0}$. The weighted MC events of $\jpsi\ar\phi\pi^{0}\pi^{0}$ are subjected to the selection criteria of  $\jpsi\ar\gamma\gamma\phi$. The surviving events are normalized according to the BF, resulting in 3155 background events which will be subtracted by giving  negative weights in the PWA. Figure~\ref{1} shows the comparison of some typical kinematic variables of data and weighted PHSP MC sample of the selected $\jpsi\ar\phi\pi^{0}\pi^{0}$ candidate events. Theses indicate good consistency between data and MC simulation. Here $\pi^{0}_{1}$ and $\pi^{0}_{2}$ denote the $\pi^{0}$s with higher and lower energies in the $\phi\piz\piz$ final state, respectively.

The surviving background events $\jpsi\ar\phi\pi^{0}$, $\jpsi\ar\phi\eta$ and $\jpsi\ar\phi\eta^{\prime}$ are normalized according to the PDG BFs~\cite{ref1}, resulting in 157 background events which also will be subtracted from the data by assigning negative weights to them in the PWA.
\section{Partial-Wave Analysis}

Using the GPUPWA framework~\cite{GPUPWA}, a PWA is performed on the selected candidate events of $\gamma\gamma\phi$, $\phi\ar\kk$, to disentangle the structures in the Dalitz plot. The amplitudes in the sequential radiative decay $\jpsi\ar\gamma R$, $R\ar\gamma\phi$ are constructed using the covariant tensor amplitudes~\cite{ref:tensor}. Let
$A_{R}$ be the amplitude of a $\jpsi$ decay process with the intermediate resonance $R$. For $\jpsi$ radiative decays, the general form of $A_{R}$ is written as 
\begin{widetext}
\begin{equation}
A_{R}(\vec{\xi})=\psi_{\mu}(p,m_{\psi})e^{*}_{\nu}(q_{1},m_{\gamma})e^{*}_{\alpha}(q_{2},m_{\gamma^{'}})A^{\mu\nu\alpha}_{R}(\vec{\xi})=\psi_{\mu}(p,m_{\psi})e^{*}_{\nu}(q_{1},m_{\gamma})e^{*}_{\alpha}(q_{2},m_{\gamma^{'}})\sum_{i}\Lambda_{R_{i}} U_{R_{i}}^{\mu\nu\alpha}(\vec{\xi}),
\label{eq4}
\end{equation}
\end{widetext}
where the sum runs over the number of independent amplitudes.
The $\psi_{\mu}(p,m_{\psi})$ is the polarization four-vector for $\jpsi$, and $e^{*}_{\nu}(q_{i},m_{\gamma})$ ($i= 1, 2$) are the polarization four-vectors for the two photons. $m$ represents the spin projection, $p$ and $q$ are the four-momenta of $\jpsi$ and photon; The $\Lambda_{R_{i}}$ is the complex coupling coefficient of the amplitude; $U_{R_{i}}^{\mu\nu\alpha}(\vec{\xi})$ is the $i$-th partial-wave amplitude for the intermediate resonance, it is constructed with the four-momenta of the particles ($\vec{\xi}$) in the final state. Their specific expressions are given in Ref.~\cite{ref:tensor}.

In this analysis, the intermediate resonances decaying into $\gamma\phi$ (denoted as $bc$) are parameterized by a constant-width BW propagator,
 \begin{equation}
f^{R}_{bc}=\frac{1}{M^{2}-s_{bc}-i\frac{1}{c^{2}}M\Gamma},
 \end{equation}
where $M$ and $\rm \Gamma$ are the mass and width of the intermediate resonance $R$, respectively; $s_{bc}$ is the invariant mass squared of $\gamma_{i}\phi$ system. The Breit-Wigner term for the non-resonant component is set to a constant in the partial-wave amplitude.

The complex coefficients of the amplitudes (relative magnitudes and phases) and the resonance parameters (masses and widths) are determined by an unbinned maximum likelihood fit to the data events. The likelihood is constructed following a method similar to the one used in Ref.~\cite{ref:gammaphiphi}.

The joint likelihood for observing $N$ events in the data sample is then given by
\begin{small}
\begin{equation}\label{joint probability density}
\mathcal{L}  =\prod\limits_{i=1}^{N}\frac{ \left|M(\vec{\xi}_{i}) \right|^2\epsilon(\vec{\xi}_{i})\Phi(\vec{\xi}_{i})  }{\sigma^\prime}.
\end{equation}
\end{small}where $\epsilon(\vec{\xi})$ is the detection efficiency, $\Phi(\vec{\xi})$ is the standard element of PHSP, $M(\vec{\xi})= \sum_{R} A_R(\vec{\xi})$ is the matrix element describing the decay $\jpsi\ar\gamma\gamma\phi$, $A_R(\vec{\xi})$ is the amplitude corresponding to the intermediate resonance $R$, and $\sigma^\prime\equiv \int{\left|M(\vec{\xi}) \right|^2\epsilon(\vec{\xi})\Phi(\vec{\xi})d\vec{\xi}}$ is the normalization integral.

Then the negative log-likelihood (NLL) value is represented by
\begin{equation}\label{minus ln L}
 -\ln \mathcal{L} = -\sum\limits_{i=1}^{N}\ln  \left|M(\vec{\xi}_{i}) \right|^2 + N \ln \sigma^\prime - \sum\limits_{i=1}^{N} \ln\epsilon(\vec{\xi}_{i})\Phi(\vec{\xi}_{i}),
\end{equation}
where the third term is a constant with no impact on the determination of the fit parameters and therefore is not considered in the fit.

The free parameters are optimized by MINUIT~\cite{MINUIT}. The normalization integral $\sigma'$ is evaluated using MC simulation with importance sampling~\cite{MCSampling1,MCSampling2}. The MC sample of size $N_{\rm gen}$ is generated with signal events uniformly distributed in PHSP. These events are simulated by taking into account of detector response, and subjected to the same selection criteria of data, yielding a sample of $N_{\rm acc}$ accepted events. The normalization integral is computed as
\begin{equation}\label{MCIntegral}
\sigma' = \int{ \left|M(\vec{\xi}) \right|^2\epsilon(\vec{\xi})\Phi(\vec{\xi})d\vec{\xi}}  \propto \frac{1}{N_{\rm gen}}\sum\limits_{j}^{N_{\rm acc}} \left|M(\vec{\xi}_{j}) \right|^2,
\end{equation}
where the constant value of the PHSP integral $\int{\Phi(\vec{\xi})d\vec{\xi}}$ is ignored.

Instead of modeling the background, the likelihood for the background is defined by the signal PDF (Eq.~\ref{joint probability density}). The contribution to the negative log-likelihood from background events in the signal region is removed by subtracting the negative log-likelihood of events in the background samples. $\mathcal{L}_{\phi\pi^{0}\pi^{0}}$ and $\mathcal{L}_{\phi\pi^{0}(\phi\eta/\phi\eta^{\prime})}$ denote the likelihoods of the $\jpsi\ar\phi\piz\piz$ events and the remaining $\jpsi\ar\phi\piz, \phi\eta, \phi\eta^{\prime}$ events, respectively.
Thus the fit minimizes the NLL given by 
\begin{widetext}
\begin{equation}\label{lilklihood}
-\ln \mathcal{L}_{\text{sig}} = -(\ln \mathcal{L}_{\text Qweighted data} - \ln \mathcal{L}_{\phi\pi^{0}\pi^{0}} -\omega_{j}\ln \cdot\mathcal{L}_{\phi\pi^{0}} -\omega_{k}\ln \cdot \mathcal{L}_{\phi\eta}
- \omega_{l}\ln \cdot \mathcal{L}_{\phi\eta^{\prime}} ).
\end{equation}
\end{widetext}
 Here, $-\ln \mathcal{L}_{\text{sig}}$ represents the negative log-likelihood of signal events, with the normalization factors $\omega_{j}=0.2$, $\omega_{k}=2.0$, $\omega_{l}=0.5$ for the $J/\psi\to \phi\pi^{0}$, $\phi\eta$, $\phi\eta^{\prime}$ background events included. The $ \mathcal{L}_{\text Qweighted data}$ and $ \mathcal{L}_{\phi\pi^{0}\pi^{0}}$ are calculated by
 
\begin{small}
\begin{equation}\label{aaa}
\mathcal{L}_{\text Qweighted data}  =\prod\limits_{i=1}^{N}\frac{ \left|M(\vec{\xi}_{i}) \right|^2\epsilon(\vec{\xi}_{i})\Phi(\vec{\xi}_{i})  }{\sigma^\prime} \cdot Q_{i},
\end{equation}
\end{small}

\begin{small}
\begin{equation}\label{bbb}
\mathcal{L}_{\phi\pi^{0}\pi^{0}}  =\prod\limits_{m=1}^{N_{\phi\pi^{0}\pi^{0}}}\frac{ \left|M(\vec{\xi}_{m}) \right|^2\epsilon(\vec{\xi}_{m})\Phi(\vec{\xi}_{m})  }{\sigma^\prime} \cdot \mathcal{W}_{m},
\end{equation}
\end{small}where $Q_{i}$ represents the Q-factor for each event, and  $\mathcal{W}_{m}$ denotes the weight for the background events of $\jpsi\ar\phi\pi^{0}\pi^{0}$. The  $\mathcal{L}_{\phi\pi^{0}}$, $\mathcal{L}_{\phi\eta}$, and $\mathcal{L}_{\phi\eta^{\prime}}$ are given as

\begin{small}
\begin{equation}\label{joint}
\mathcal{L}_{\phi\pi^{0} (\phi\eta/\phi\eta^{\prime})}  =\prod\limits_{i=1}^{N_{\phi\pi^{0} (\phi\eta / \phi\eta^{\prime})}}\frac{ \left|M(\vec{\xi}_{i}) \right|^2\epsilon(\vec{\xi}_{i})\Phi(\vec{\xi}_{i})  }{\sigma^\prime}.
\end{equation}
\end{small}

The log-likelihood is scaled by taking into account the statistical uncertainties of subtracted backgrounds~\cite{scale}.

The fitted event yield $N_{R}$ of an intermediate resonance $R$ is determined as
\begin{eqnarray}\label{13}
N_R = \frac{\sigma_R}{\sigma'}\cdot N_{\rm sig},
\end{eqnarray}
where $N_{\rm sig}$ the number of signal events after subtracting all backgrounds, and
\begin{eqnarray}\label{14}
\sigma_R = \frac{1}{N_{\rm gen}}\sum_{j}^{N_{\rm acc}}|A_R(\vec{\xi}_j)|^2
\end{eqnarray}
is calculated with the same MC sample as the normalization integral $\sigma'$.

The interference cross section between two different resonances is calculated by
\begin{equation}
\sigma_{R_1,R_2} = \frac{1}{N_{acc}}\sum_{k}^{N_{acc}}\sum_{j_1}^{N_{W_1}}\sum_{j_2}^{N_{W_2}}((A_{j_1})_k(A^{*}_{j_2})_k + (A^{*}_{j_1})_k(A_{j_2})_k),
\end{equation}
where $N_{W_1}$ and $N_{W_2}$ are the numbers of the independent partial-wave amplitudes of $R_1$ and $R_2$, respectively. 

The detection efficiency $\epsilon_R$ of the intermediate resonance $R$ is obtained by the partial-wave amplitude weighted MC sample,
\begin{eqnarray}\label{19}
\epsilon_R =  \frac{\sum_{j}^{N_{\rm acc}}|A_R(\vec{\xi}_j)|^2}{\sum_{n}^{N_{\rm gen}}|A_R(\vec{\xi}_n)|^2}~.
\end{eqnarray}
Finally, the BF of $J/\psi\ar\gamma R, R\ar\gamma\phi$ is calculated by
\begin{equation}
\mathcal{B}(J/\psi\ar\gamma R \ar \gamma\gamma\phi) = \frac{N_{R}}{N_{J/\psi}\cdot \epsilon_R \cdot \mathcal{B}(\phi\ar\kk)},
\label{BR}
\end{equation}
where $N_{J/\psi}$ is the total number of $J/\psi$ event, which is $(10087\pm44)\times 10^{-6}$; $\epsilon_R$ is the detection efficiency of the resonance $R$; $\mathcal{B}(\phi\ar\kk)=(49.1\pm0.5)\%$ is the BF of $\phi\ar\kk$ taken from the
PDG~\cite{ref1}. 

\section{Analysis Results}
In this analysis, we consider all kinematically-allowed resonances with $J^{PC} = 0^{++}$, $0^{-+}$, $1^{++}$, $1^{-+}$, $2^{++}$ and $2^{-+}$ as listed from the PDG~\cite{ref1}. Additionally, we take into account an exotic structure $\eta_{1}(1855)$ recently observed in $\jpsi\ar\gamma\eta\eta^{\prime}$ by BESIII~\cite{1855,1855_2} and a glueball candidate $X(2370)$ observed in $\jpsi\ar\gamma\pi^+\pi^-\eta^{\prime}$~\cite{ref11}. We also consider non-resonant contributions with certain $J^{PC} = 0^{++}$, $0^{-+}$, $1^{++}$, $1^{-+}$, $2^{++}$ for the $\gamma\phi$ system. All possible resonances listed in Table~\ref{Candidate poor} are taken into account, and their statistical significances are evaluated. Only the components with statistical significance greater than 5$\sigma$ are kept in the baseline solution. The statistical significance of each resonance is determined 
from the relative changes of optimal NLL values when including and excluding it in the fit, assuming the $\chi^{2}$ distribution hypothesis and taking into account the change in the number of degrees of freedom.
\begin{table}[!htp]
\caption{Pool of candidate resonances for the PWA.}
\begin{center}
\label{Candidate poor}
\begin{tabular}{c|c|c|c|c|c} \hline\hline
$0^{++}$ &$0^{-+}$ &$1^{++}$ &$1^{-+}$&$2^{++}$ &$2^{-+}$  \\ \hline
$f_{0}(1370)$ &$\eta(1295)$ &$f_{1}(1285)$& $\eta_{1}(1855)$&$f_{2}(1270)$ &$\eta_{2}(1645)$ \\
$f_{0}(1500)$ &$\eta(1405)$ &$f_{1}(1420)$&                 &$f_{2}(1430)$ &$\eta_{2}(1870)$ \\
$f_{0}(1710)$ &$\eta(1475)$ &$f_{1}(1510)$&                 &$f_{2}(1525)$ &                \\
$f_{0}(2020)$ &$\eta(1760)$ &             &                 &$f_{2}(1565)$ &                 \\
$f_{0}(2100)$ &$X(1835)$    &             &                 &$f_{2}(1640)$ &                 \\
$f_{0}(2200)$ &$\eta(2225)$ &             &                 &$f_{2}(1810)$ &                 \\
$f_{0}(2330)$ &$X(2370)$    &             &                 &$f_{2}(1910)$ &                 \\
              &$\eta_{c}$   &             &                 &$f_{2}(1950)$ &                 \\
              &             &             &                 &$f_{2}(2010)$ &                 \\
              &             &             &                 &$f_{2}(2150)$ &                 \\
              &             &             &                 &$f_{J}(2220)$ &                   \\
              &             &             &                 &$f_{2}(2300)$ &                   \\
              &             &             &                 &$f_{2}(2340)$ &                   \\
\hline\hline
\end{tabular}
\end{center}
\end{table}

\begin{figure*}[!htp]
\begin{center}
\includegraphics[width=0.45\textwidth]{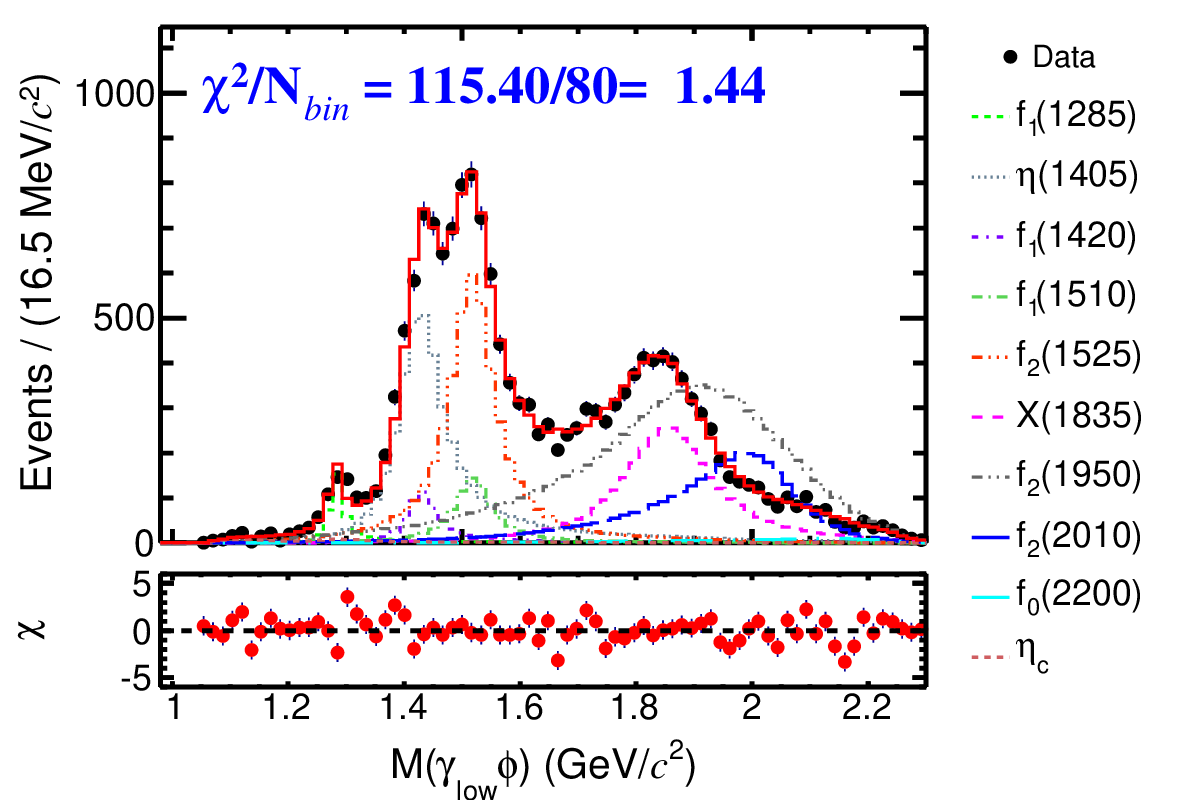}
\put(-70,110){\textbf{(a)}}
\includegraphics[width=0.45\textwidth]{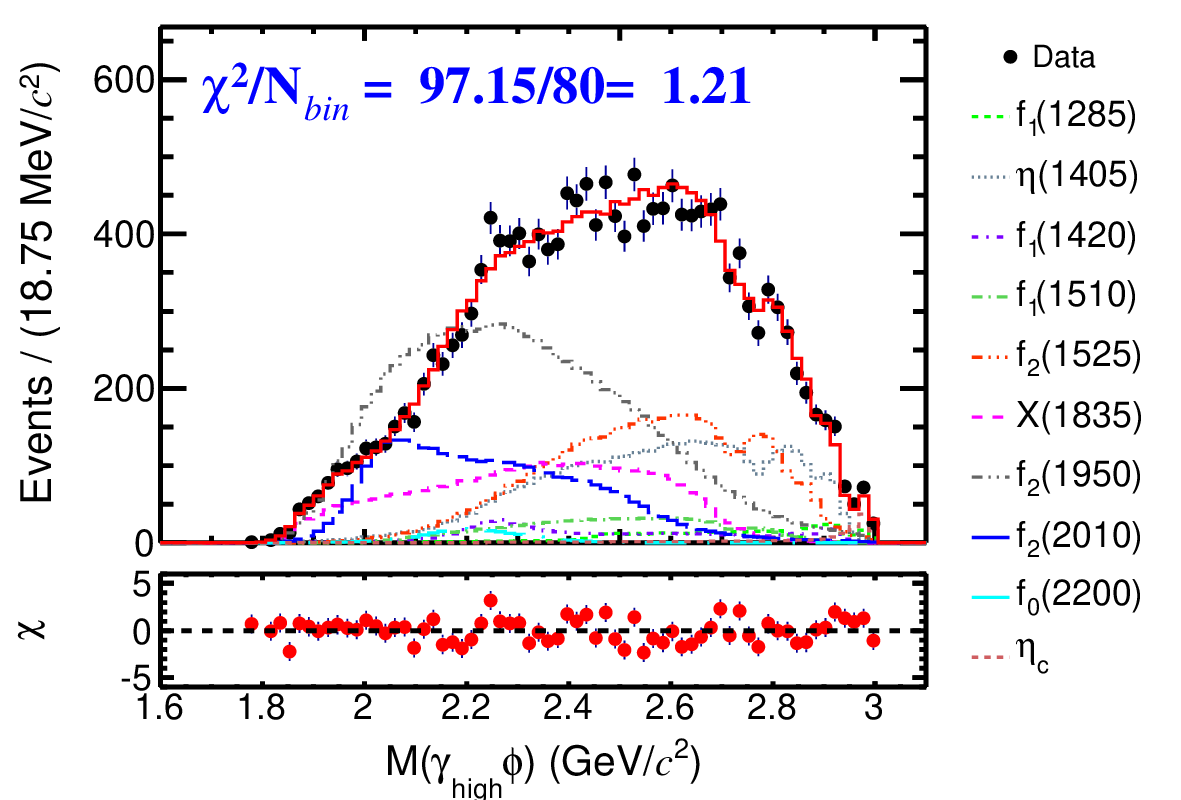}
\put(-70,110){\textbf{(b)}}

\includegraphics[width=0.45\textwidth]{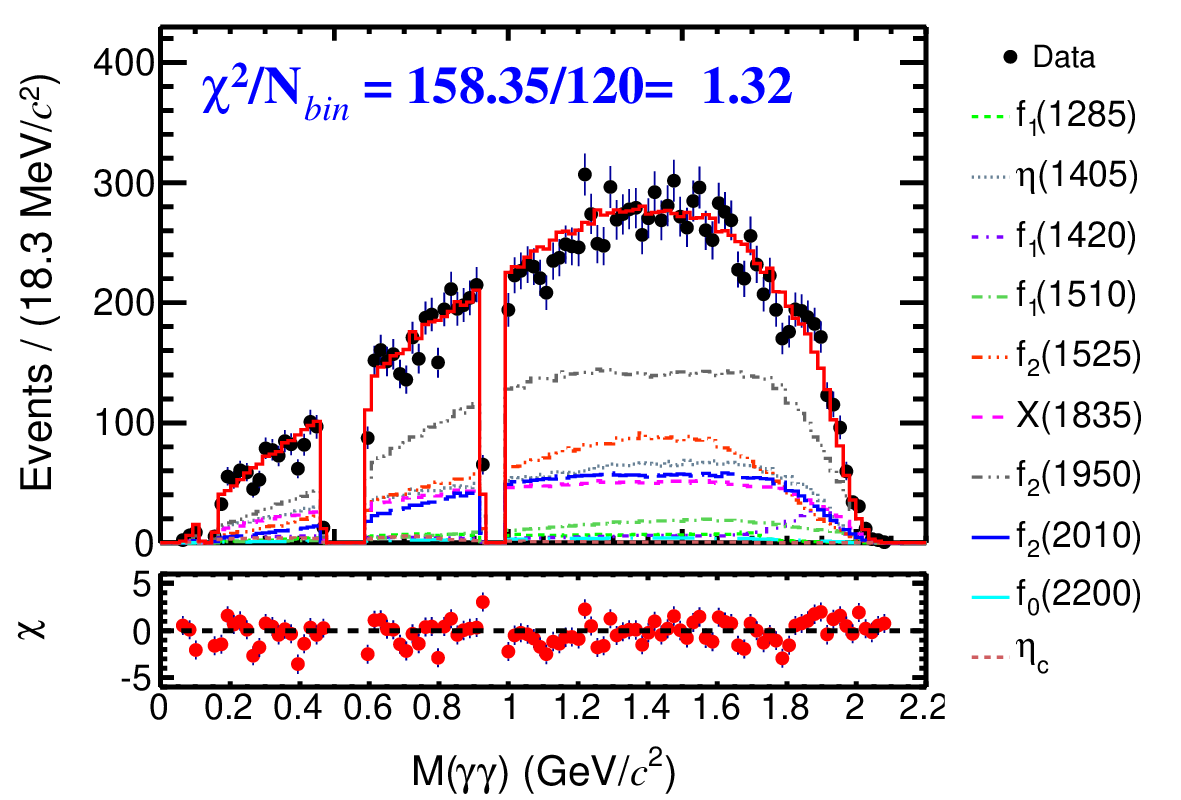}
\put(-70,117){\textbf{(c)}}
\includegraphics[width=0.45\textwidth]{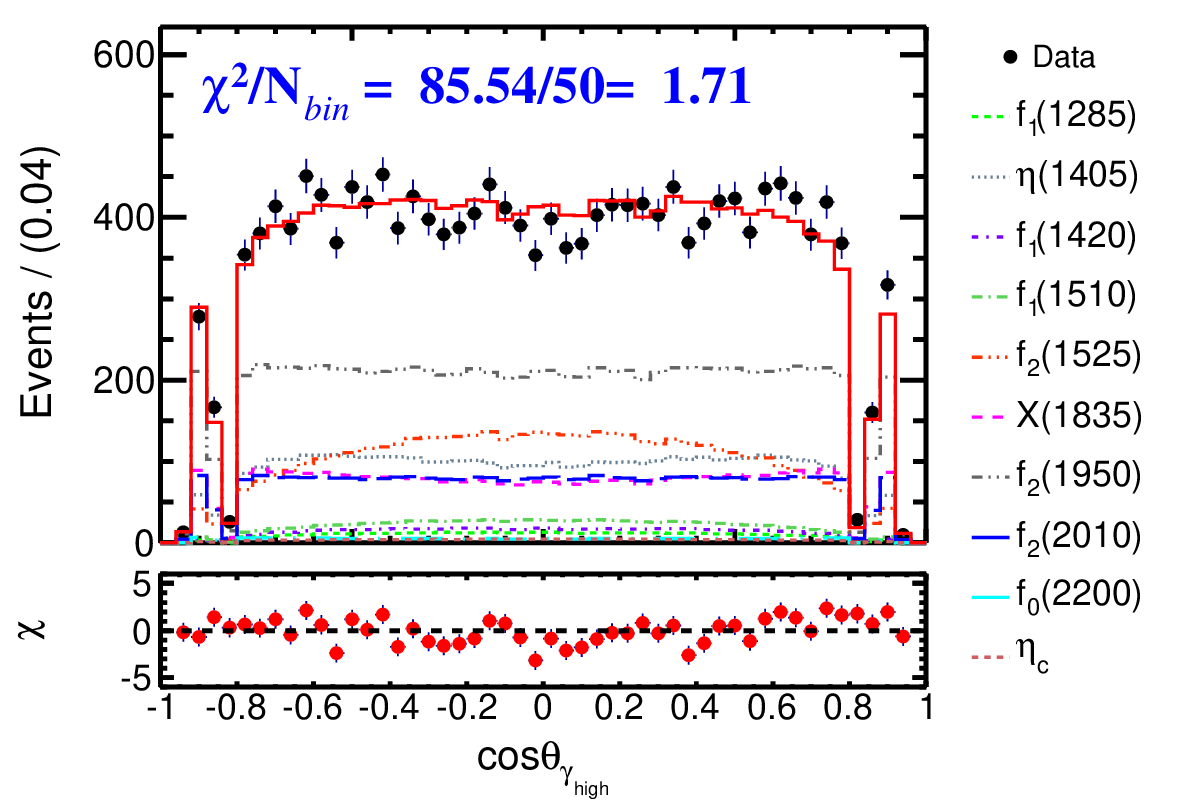}
\put(-70,117){\textbf{(d)}}

\includegraphics[width=0.45\textwidth]{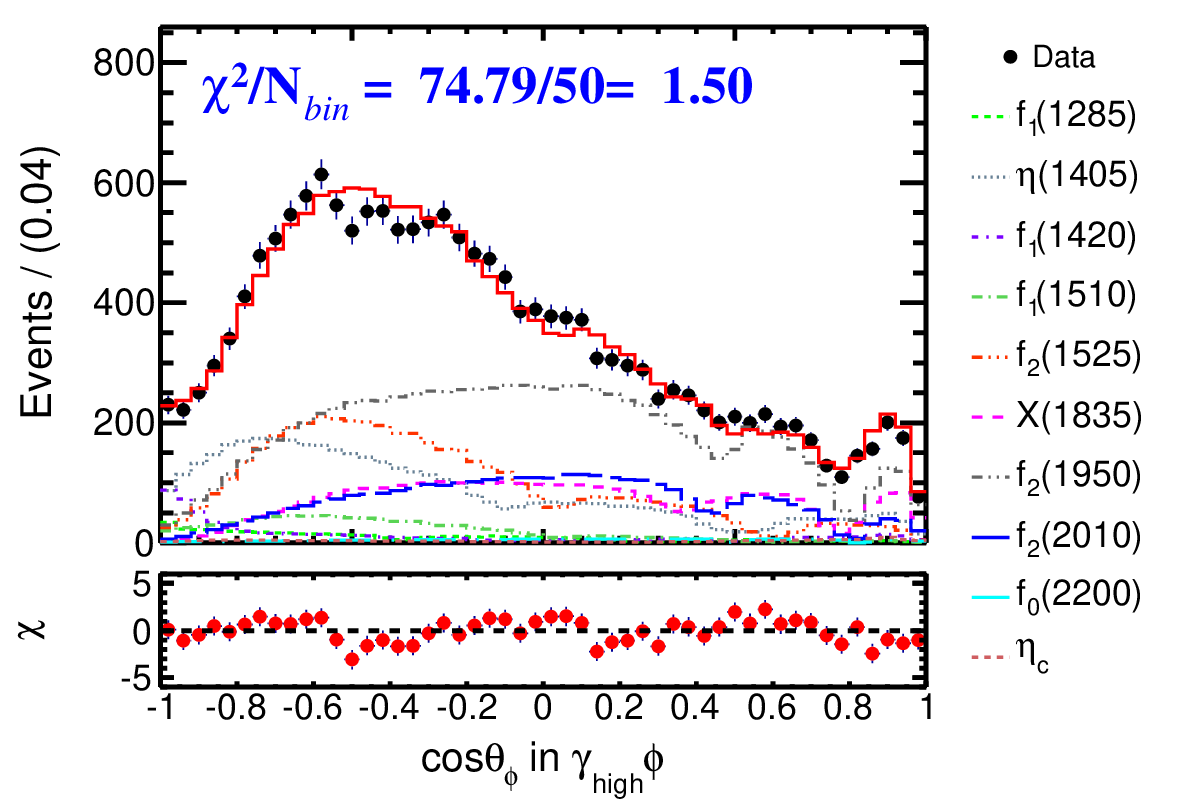}
\put(-70,110){\textbf{(e)}}
\includegraphics[width=0.45\textwidth]{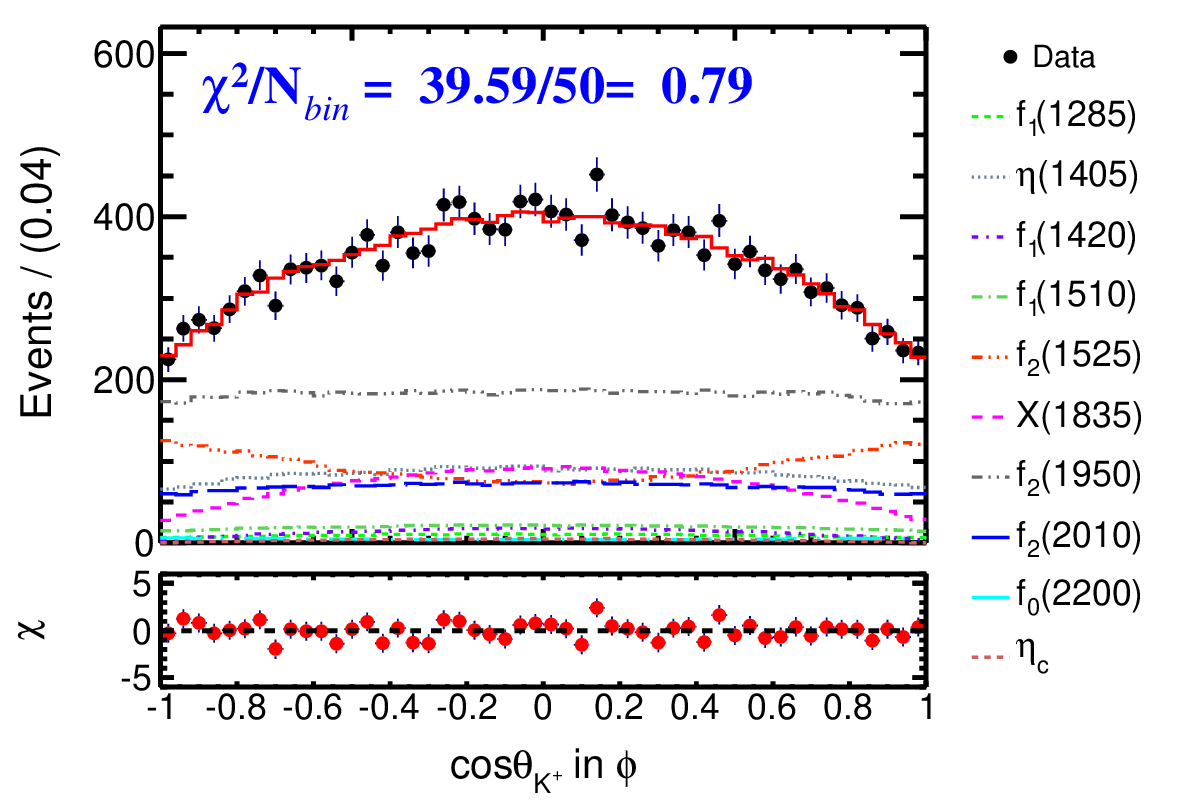}
\put(-70,110){\textbf{(f)}}

\includegraphics[width=0.45\textwidth]{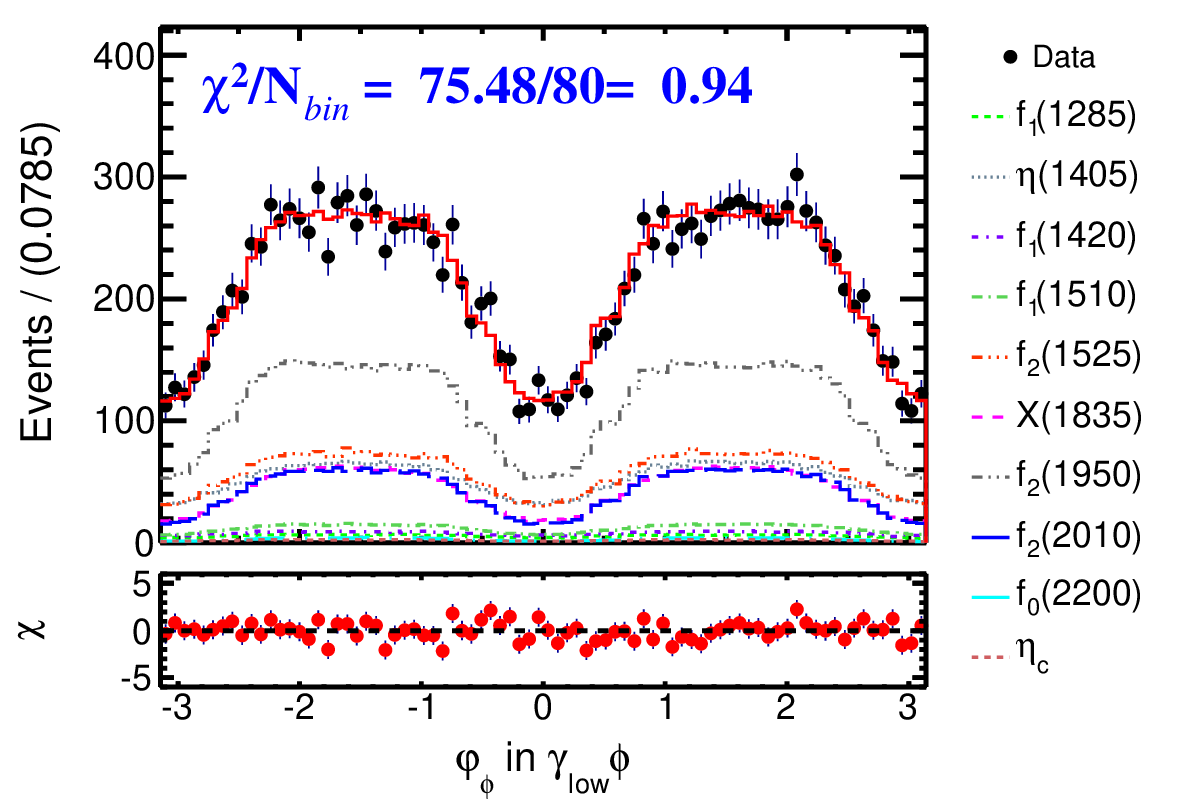}
\put(-70,120){\textbf{(g)}}
\includegraphics[width=0.45\textwidth]{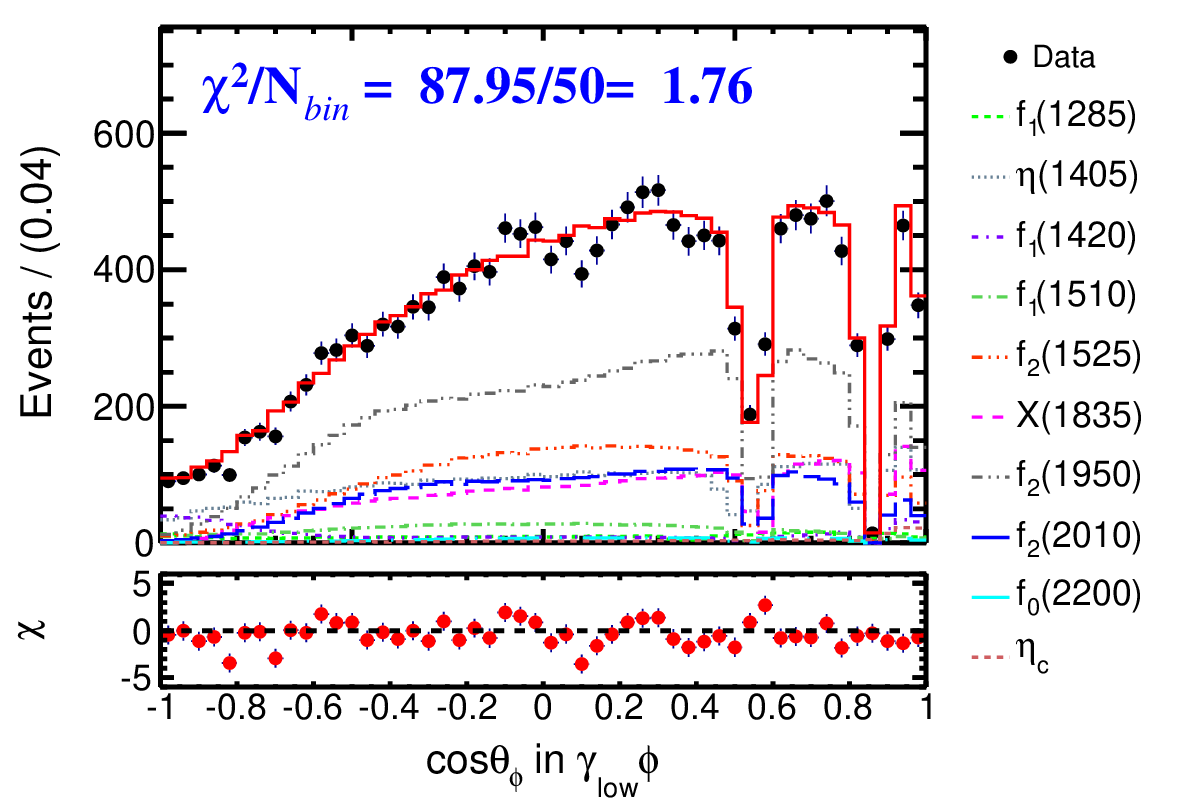}
\put(-70,120){\textbf{(h)}}
\renewcommand{\figurename}{FIG.}
\caption{PWA fit projections on (a) $M(\gamma_{\text low}\phi)$, (b) $M(\gamma_{\text high}\phi)$, (c) $M(\gamma\gamma)$, (d) cosine of polar angle of the high-energy photon in the rest frame of $\jpsi$, (e) cosine of polar angle of $\phi$ in the rest frame of $\gamma_{\text high}\phi$, (f) cosine of polar angle of $K^{+}$ in the rest frame of $\phi$, (g) azimuth of $\phi$ in the rest frame of $\gamma_{\text low}\phi$, and (h) cosine of polar angle of $\phi$ in the rest frame of $\gamma_{\text low}\phi$. The black dots with error bars represent data, the red lines represent the projections of PWA fit, and the bashed lines represent contributions of each component in the baseline solution. }
\label{pro}
\end{center}
\end{figure*}

Due to limited statistics, the masses, widths and $J^{PC}$ of all candidate states are fixed to individual PDG values. However, for $\eta(1405)$ and $X(1835)$, their masses and widths are free. Changing the $J^{PC}$ in the baseline set of amplitudes to other hypotheses results in a worse negative log-likelihood by at least 13.2 units. 
The results of the baseline fit including the masses, widths and product BFs are listed in Table~\ref{basic}. 
The baseline results show that the structure in the vicinity of 1.4 GeV/$c^{2}$, as shown in Fig.~\ref{pro}(a), needs to be described by $\eta(1405)$ and $f_{1}(1420)$ in the $\gamma\phi$ system. The measured mass and width of $\eta(1405)$ are $(1422.0 \pm 2.1 ^{+5.9}_{-7.8})$~MeV/$c^{2}$ and $(86.3 \pm 2.7 ^{+6.6}_{-17.4})$~MeV, respectively. The spin-parity of the structure with mass around 1.8 GeV/$c^{2}$ is determined to be $0^{-+}$. Its mass and width are measured to be $(1849.3\pm3.0^{+7.6}_{-10.0})$ MeV/$c^{2}$ and $(179.6 \pm 8.7 ^{+22.5}_{-27.9})$~ MeV, which are consistent with those of the $X(1835)$ measured in $\jpsi\ar\gamma K^{0}_{S}K^{0}_{S}\eta$~\cite{ref5}. 
The fit fractions of each component (defined as $\sigma_{R}/\sigma^{\prime}$) and their interference fractions (defined as $\sigma_{R_{1},R_{2}}/\sigma^{\prime}$) are shown in Table~\ref{interference}. 
The comparisons between the data and the PWA fit projections 
for the distributions of $M(\gamma_{\text low}\phi)$, $M(\gamma_{\text high}\phi)$, $M(\gamma\gamma)$ and several angular distributions are shown in Fig.~\ref{pro}. The $\chi^{2}/N_{{\rm bin}}$ is displayed on each
figure to demonstrate the goodness of fit, where $N_{{\rm bin}}$ is the number of bins in each histogram, and 
$\chi^{2}$ is defined as
\begin{eqnarray}
\chi^{2}=\sum_{i=1}^{N_{{\rm bin}}}\frac{(n_{i}-\nu_{i})^{2}}{\nu_{i}},
\end{eqnarray}
where $n_{i}$ and $\nu_{i}$ are the numbers of events of the data and
the fit projections with the baseline set of amplitudes in the
$i$th bin of each figure, respectively.

\begin{table*}[!htp]
\caption{The masses and widths of resonances, significance and the product
BFs $ \mathcal{B}(\jpsi\ar\gamma R\ar\gamma\gamma\phi)$ of each component in the PWA fit using the baseline set of amplitudes. The first uncertainties are statistical and the second are systematic. The masses and widths of the $f$-states and $\eta_{c}$ are fixed to the PDG values~\cite{ref1}.}
\begin{center}
\label{basic}
\begin{tabular}{c|cccc} \hline\hline
Resonance    & $M$ (MeV/$c^2$)     &$\Gamma$ (MeV)                   & $ \mathcal{B}~$($\times10^{-6}$)    &Significance\\ \hline
$f_{1}(1285)$ & 1281.9 &22.7 &  0.29$\pm$0.03$^{+0.11}_{-0.09}$          &17.3$\sigma$\\
$f_{1}(1420)$& 1426.3   &54.5& 0.55$\pm$0.07$^{+0.18}_{-0.17}$   &9.0$\sigma$\\
$\eta(1405)$ & $1422.0\pm2.1^{+5.9}_{-7.8}$  &$86.3\pm2.7^{+6.6}_{-17.4}$   & 3.57$\pm$0.18$^{+0.59}_{-0.61}$          &18.9$\sigma$\\
$f_{1}(1510)$& 1518.0              & 73.0                              & 0.78$\pm$0.09$^{+0.34}_{-0.30}$          &5.3$\sigma$\\
$f_{2}(1525)$& 1517.4            &86.0                               & 2.76$\pm$0.18$^{+0.90}_{-0.61}$          &16.4$\sigma$\\
$X(1835)$   & $1849.3\pm3.0^{+7.6}_{-10.0}$  &$179.6\pm8.7^{+22.5}_{-27.9}$                      & 3.37$\pm$0.19$^{+0.78}_{-1.10}$          &15.3$\sigma$\\
$f_{2}(1950)$& 1936.0              &464.0                              & 9.96$\pm$0.60$^{+3.44}_{-2.13}$          &13.1$\sigma$\\
$f_{2}(2010)$& 2011.0              &202.0                              & 4.63$\pm$0.43$^{+1.42}_{-1.46}$          &11.3$\sigma$\\
$f_{0}(2200)$& 2187.0              &207.0                              & 0.20$\pm$0.04$^{+0.05}_{-0.07}$          &6.3$\sigma$\\
$\eta_{c}$   & 2983.9            &32.0                               & 0.21$\pm$0.03$^{+0.05}_{-0.07}$          &12.9$\sigma$\\ \hline\hline
\end{tabular}
\end{center}
\end{table*}

\begin{table*}[!htp]

\caption{Fit fractions ($\sigma_{R}/\sigma^{\prime}$) of each component and interference fractions ($\sigma_{R_{1},R_{2}}/\sigma^{\prime}$) between two components in the baseline solution. The uncertainties are statistical only. All the values are given in \%.}
\begin{center}
\label{interference}
\begin{tabular}{c|cccccccccc} \hline\hline
Resonance     & $f_{1}(1285)$  & $f_{1}(1420)$ & $\eta(1405)$   &  $f_{1}(1510)$ & $f_{2}(1525)$  &  $X(1835)$     & $f_{2}(1950)$  & $f_{2}(2010)$  &$f_{0}(2200)$   & $\eta_{c}$ \\ \hline

$f_{1}(1285)$ & 1.84$\pm$0.16 & 0.54$\pm$0.11 & 0.34$\pm$0.08 &-0.26$\pm$0.11 &-0.31$\pm$0.04  &-0.34$\pm$0.09 &-0.14$\pm$0.07  &-0.09$\pm$0.05 &0.00$\pm$0.00 & 0.07$\pm$0.03  \\ 
$f_{1}(1420)$ &                & 3.94$\pm$0.42 & -0.30$\pm$0.21 &-2.38$\pm$0.45  &-0.16$\pm$0.22  &1.00$\pm$0.18  & -3.42$\pm$0.32& 2.45$\pm$0.22 &0.00$\pm$0.00  & -0.18$\pm$0.05   \\ 
$\eta(1405)$  &                &               & 21.85$\pm$0.79 &-0.78$\pm$0.17  &-0.29$\pm$0.09  &2.14$\pm$0.82   &-0.19$\pm$0.09  &-0.14$\pm$0.05 &0.02$\pm$0.01 & 1.66$\pm$0.16   \\ 
$f_{1}(1510)$ &                &               &                &5.62$\pm$0.56  &0.55$\pm$0.42   &0.17$\pm$0.16   &4.85$\pm$0.45  &-3.10$\pm$0.27 &-0.13$\pm$0.05  & -0.13$\pm$0.05   \\ 
$f_{2}(1525)$ &                &               &                &                &18.74$\pm$0.95  &0.09$\pm$0.06   &7.83$\pm$0.96   &-2.14$\pm$0.45  & 0.18$\pm$0.06  & -0.12$\pm$0.01   \\ 
$X(1835)$     &                &               &                &                &                &19.03$\pm$0.76  &1.88$\pm$0.17  &-1.12$\pm$10.13&0.00$\pm$0.00 & -2.13$\pm$0.14 \\ 
$f_{2}(1950)$ &                &               &                &                &                &                &60.94$\pm$2.78  &-72.11$\pm$4.56 &-1.06$\pm$0.21  & 0.24$\pm$0.04  \\ 
$f_{2}(2010)$ &                &               &                &                &                &                &                &19.03$\pm$0.76  &1.85$\pm$0.26  & -0.14$\pm$0.02 \\ 
$f_{0}(2200)$ &                &               &                &                &                &                &                &                &30.78$\pm$2.45  & 0.01$\pm$0.00 \\ 
$\eta_{c}$    &                &               &                &                &                &                &                &                &                & 1.28$\pm$0.14 \\ \hline\hline
\end{tabular}
\end{center}
\end{table*}

All other possible resonances have a statistical significance less than $5\sigma$ when added to the baseline set of amplitudes with fixed resonance parameters as listed in Table~\ref{basic}. In addition, the statistical significances of all possible non-resonant contributions are also less than 5$\sigma$, as shown in Table~\ref{extra}. Among them, the $f_{0}(2020)$ has a significance of 4.9$\sigma$, but with a fit fraction much less than 1\%. In the following text, the upper limit on the BF of $\jpsi\ar\gamma f_{0}(2020)\ar\gamma\gamma\phi$ is presented at 90\% confidence level.
 The impact of the inclusion of $f_{0}(2020)$ on the PWA result is assigned as a systematic uncertainty, as discussed in Sec.~VI. To investigate additional possible contributions, resonances with different $J^{PC}$ ($0^{++}$, $0^{-+}$, $1^{++}$, $1^{-+}$, $2^{++}$, $2^{-+}$) and different masses and widths are added to the baseline set of amplitudes. No significant contribution is found.
 
 \begin{table}[!htp]
\caption{Changes in the negative log-likelihood ($\Delta \ln \mathcal{L}$) and in the number of degrees of freedom ($\Delta$dof) by considering additional resonances and the corresponding statistical significance. The $\Delta$dof is determined by the complex coefficient of independent partial wave~\cite{ref:tensor}.}
\begin{center}
\label{extra}
\begin{tabular}{l r@{.}l rc} \hline\hline
Resonance   &\multicolumn{2}{l}{$\Delta \ln \mathcal{L}$}   & $\Delta$dof &Significance \\ \hline

$f_{2}(1270)$   &  3&9    &  18    &0.1$\sigma$\\
$\eta(1295)$    &  0&1    &  2     &0.1$\sigma$\\
$f_{0}(1370)$   &  1&4    &2   &1.2$\sigma$\\
$f_{0}(1430)$   &  1&4   &2        &1.2$\sigma$\\ 
$f_{2}(1430)$   &  4&9   &18        &0.1$\sigma$\\ 
$\eta(1475)$    & 8&6     &  2     &3.7$\sigma$\\
$f_{0}(1500)$   & 0&8     &  2     &0.7$\sigma$\\
$f_{2}(1565)$   &  12&4   &  18    &1.5$\sigma$\\
$f_{2}(1640)$   &  7&4    &  18    &0.4$\sigma$\\
$\eta_{2}(1645)$&  3&3    &  18    &0.0$\sigma$\\
$f_{0}(1710)$   &  2&7    &  2     &1.8$\sigma$\\
$\eta(1760)$    &  0&6    &  2     &0.6$\sigma$\\
$f_{2}(1810)$   &  10&3   &  18    &1.0$\sigma$\\
$\eta_{1}(1855)$&  2&9    &  8     &0.4$\sigma$\\
$\eta_{2}(1870)$&  11&6    &  18     &1.3$\sigma$\\
$f_{2}(1910)$   &  9&6    &  18    &0.9$\sigma$\\
$f_{0}(2020)$   &  13&6    &  2     &4.9$\sigma$\\
$f_{0}(2100)$   &  7&4    &  2     &3.4$\sigma$\\
$f_{2}(2150)$   &  9&6    &  18    &0.9$\sigma$\\
$f_{J}(2220)$   &  13&7    &  18    &1.8$\sigma$\\
$\eta(2225)$    &  0&6    &  2     &0.6$\sigma$\\
$f_{2}(2300)$   &  16&8    &  18     &2.5$\sigma$\\
$f_{0}(2330)$   &  12&0    &  2     &4.5$\sigma$\\
$f_{2}(2340)$   &  15&2    &  18    &2.1$\sigma$\\
$X(2370)$       &  0&4     &  2     &0.4$\sigma$\\
$0^{++}$ PHSP   &  0&9    &  2     &0.9$\sigma$\\
$0^{-+}$ PHSP   &  4&1    &  2     &2.4$\sigma$\\
$1^{++}$ PHSP   &  5&9   &  8     &1.5$\sigma$\\
$2^{++}$ PHSP   &  18&4   &  18    &2.8$\sigma$\\
$2^{-+}$ PHSP   &  5&9    &  18    &0.2$\sigma$\\
\hline\hline

\end{tabular}
\end{center}
\end{table}

A further check on the pseudoscalar structure with mass around 1.4 GeV/$c^2$ is performed. If the mass and width of the pseudoscalar state are fixed to the PDG values of $\eta(1405)$ or $\eta(1475)$~\cite{ref1}, the log-likelihood worsens by 45.6 or 56.2. Another test includes an additional $\eta(1475)$ with parameters fixed at the PDG values or free resonance parameters in the baseline model, the statistical significances of these contributions are 2.3$\sigma$ and 3.9$\sigma$. These results indicate that the possibility of a second $\eta$ state around 1.4 GeV/$c^2$ is not excluded.

The PWA result shows that there is no evidence for the $\eta(1295)$, $\eta(1475)$, $\eta_{1}(1855)$, $f_{0}(2020)$ and X($2370)$ in the $\gamma\phi$ system. The same approach as in Ref.~\cite{ref261} is used to determine the upper limits of the production of these five states. The systematic uncertainties of these upper limits are calculated with same method as for the baseline solution, cf. Sec.~\ref{sec:sysunc}, and are summed in quadrature with the statistical uncertainty for each respective yield. The resulting uncertainty is used to determine the 90\% confidence level deviation, and added to the updated nominal yield of the candidate state to obtain the corresponding upper limit on the BF. As a result, the upper limits are determined to be $ \mathcal{B}(\jpsi\ar\gamma\eta(1295)\ar\gamma\gamma\phi)<8.37\times10^{-7}$, $ \mathcal{B}(\jpsi\ar\gamma\eta(1475)\ar\gamma\gamma\phi)<3.80\times10^{-7}$, $ \mathcal{B}(\jpsi\ar\gamma\eta_{1}(1855)\ar\gamma\gamma\phi)<4.74\times 10^{-6}$, $\mathcal{B}(\jpsi\ar\gamma f_{0}(2020)\ar\gamma\gamma\phi)<2.41\times 10^{-6}$ and $ \mathcal{B}(\jpsi\ar\gamma X(2370)\ar\gamma\gamma\phi)<1.08\times 10^{-7}$.


\section{Systematic Uncertainties}
\label{sec:sysunc}
\subsection{Systematic uncertainties due to event selection}
The systematic uncertainties that arise from event selection impact the BF measurements. The sources contributing to the uncertainty are listed below.
\begin{itemize}
\item[(i)] Total number of $\jpsi$ events. The uncertainty due to the total number of $\jpsi$ events is 0.4\% according to Ref.~\cite{Njpsi}.
\item[(ii)] Kaon tracking and PID. The charged kaon tracking and PID efficiencies are studied using control sample $\jpsi\ar K^{0}_{S}K^{\pm}\pi^{\mp}$, $K^{0}_{S}\ar\pi^+\pi^-$~\cite{ref28}. The systematic uncertainty from tracking and PID is estimated to be 1\% per kaon.
\item[(iii)] Photon detection. The photon detection efficiency is studied using $\jpsi\ar\rho\pi^{0}\ar\pi^+\pi^-\pi^0$~\cite{ref29} control sample. The systematic uncertainty of photon detection is assigned to be 1\% per photon.
\item[(iv)] Kinematic fit. The track helix parameter correction method~\cite{ref30} is used to investigate the systematic uncertainty associated with the 4C kinematic fit. The difference in the detection efficiencies with and without the helix parameter correction, 1.2\%, is taken as the systematic uncertainty.
\end{itemize}
The total systematic uncertainty from event selection is determined to be 3.5$\%$, obtained as a quadratic sum of the contributions discussed above. 

\subsection{Systematic uncertainties due to PWA}
Systematic uncertainty from the PWA impacts both BFs and resonance parameters. The sources of this uncertainty are described below.
\begin{itemize}
\item[(i)] Mass window cuts: The uncertainties from mass window requirements of $\chicz$ are estimated by enlarging or shrinking each mass window within 2 MeV/$c^{2}$. The largest changes in the measured masses and widths of $\eta(1405)$, $X(1835)$ and the BFs are assigned as the systematic uncertainties.
\item[(ii)] Non-$\phi$ backgrounds: The uncertainty due to the $Q$-factor method~\cite{qfactor} is obtained by smearing the nominal $Q$-value in each event
by a Gaussian distribution with standard deviation 
\begin{equation}
\sigma^2_Q = \sum_{ij}\frac{\partial Q}{\partial \alpha_{i}}(C^{-1}_{\alpha})_{ij}\frac{\partial Q}{\alpha_{j}},
\end{equation}
where $Q$ is the $Q$-factor, $\alpha_{i}$ are the measurable quantities and $C_{\alpha}$ is the full covariance matrix. The new $Q$-factors are used in the PWA, and the differences between the nominal and new fits are assigned as the systematic uncertainties.
\item[(iii)] $\phi$ backgrounds: The ``data-like” MC distributions of $\jpsi\ar\phi\pi^0\pi^0$ have been obtained using a machine learning-based multi-dimensional reweighting method~\cite{reweight}. The systematic uncertainties are estimated by
changing the kinematic variables utilized for machine learning training.
 The kinematic variables are changed from four-momenta of $\pi^{0}_{1}$, $\pi^{0}_{2}$ and $\phi$, $M(\phi\pi^{0}_{1})$,  $M(\phi\pi^{0}_{2})$,  $M(\pi^{0}_{1}\pi^{0}_{2})$, cos$\theta_{\phi}$, cos$\theta_{\pi^{0}_{1}}$, and cos$\theta_{\pi^{0}_{2}}$ to $P_{z}(\pi_{1}^{0})$, $E(\pi_{1}^{0})$, $P_{z}(\pi_{2}^{0})$, $E(\pi_{2}^{0})$, $P_{z}(\phi)$, $E(\phi)$, $M(\phi\pi^{0}_{1})$, $M(\phi\pi^{0}_{2})$, $M(\pi^{0}_{1}\pi^{0}_{2})$, $\varphi_{\phi}$, $\varphi_{\pi^{0}_{1}}$, and $\varphi_{\pi^{0}_{2}}$. The changes in the results are regarded as the systematic uncertainties. Since the remaining contributions of $\jpsi\ar\phi \pi^{0}$, $\jpsi\ar\phi\eta$, and $\jpsi\ar\phi\eta^{\prime}$ are small~($\sim$0.8\%), their impact on the PWA results is negligible.
\item[(iv)] Resonance parameters: In the baseline solution, the resonance parameters of $f_1(1285)$, $f_1(1420)$, $f_2(2010)$, $f_0(2200)$, $f_1(1510)$, $f_2(1950)$, $f_2(1525)$, and $\eta_{c}$ are fixed to the PDG~\cite{ref1} values. Alternative fits are performed with these resonance parameters varied within one standard deviation of individual PDG value. The changes in the results are taken as the corresponding systematic uncertainties.
\item[(v)] Additional resonances: The uncertainties arising from possible additional resonances are estimated by adding the $f_{0}(2020)$, $\eta(1405)$, $1^{++}$ PHSP, $2^{++}$ PHSP, or $\eta_{2}(1870)$ into the baseline fit individually. These candidate resonances are the most significant additional resonances for each possible $J^{PC}$. The resulting changes in the measurements are assigned as the systematic uncertainties.
\end{itemize}

For each measurement, the individual uncertainties are assumed to be independent and are added in quadrature to obtain 
the total systematic uncertainty. The sources of systematic uncertainties in the measurements of masses and widths of $\eta(1405)$ and $X(1835)$ are summarized in Table~\ref{sum1}. The relative systematic uncertainties of the BF measurements are summarized in Table~\ref{sum2}.

\begin{table*}[!htp]
\caption{Systematic uncertainties on the masses and widths of $\eta(1405)$ and $X(1835)$.}
\begin{center}
\label{sum1}
\renewcommand\arraystretch{1.2}
\begin{tabular}{c|cc|cc}\hline\hline
\multirow{2}{*}{Source} &    \multicolumn{2}{c|}{$\eta(1405)$} &  \multicolumn{2}{c}{$X(1835)$} \\
\cline{2-5} &$\Delta M$ (MeV$/c^{2}$)&   $\Delta \Gamma$ (MeV) & $\Delta M $(MeV$/c^{2}$) &   $\Delta \Gamma$ (MeV) \\ \hline
$\pi^{0}$ mass window cut    & 2.5  & 2.7   & 3.1& 9.4 \\  
$\eta$ mass window cut   &  2.0  &3.0   &1.0  & 8.8 \\ 
$\eta^{\prime}$ mass window cut   & 1.6& 1.9 & 0.5 & 2.6 \\ 
$Q$-factor method &2.0 &1.9 &0.4 &9.4 \\ 
Multi-dimensional reweighting method&2.5  &2.7 &3.1 &8.9  \\  
Resonance parameters& 2.0 & 2.7&3.0 &8.6  \\ 
Extra resonances&$^{+2.8}_{-5.8}$  &$^{+2.2}_{-16.3}$  &$^{+5.4}_{-8.4}$  & $^{+9.7}_{-19.1}$   \\ \hline
Total & $^{+5.9}_{-7.8}$ &$^{+6.6}_{-17.4}$ &$^{+7.7}_{-10.0}$  &$^{+22.5}_{-27.9}$\\ 
 \hline \hline
\end{tabular}
\end{center}
\end{table*}

\begin{table*}[!htp]
\caption{Relative systematic uncertainties on $\mathcal{B}(\jpsi\ar\gamma R \ar\gamma\gamma\phi)~(\%)$. }
\begin{center}
\label{sum2}
\renewcommand\arraystretch{1.5}
\begin{tabular}{c|cccccccccc}\hline\hline
Source &$f_{1}(1285)$  & $f_{1}(1420)$ & $f_{0}(2200)$  & $f_{2}(2010)$&  $\eta(2983)$    & $\eta(1405)$  & $f_{1}(1510)$  & $X(1835)$    &$f_{2}(1950)$ & $f_{2}(1525)$ \\ \hline
Event selection & 3.5 & 3.5 & 3.5 & 3.5 & 3.5 & 3.5  &3.5  & 3.5& 3.5 &3.5 \\ 
$\pi^{0}$ mass window cut & 10.7 & 6.6 & 6.3 & 13.7 & 11.8 &6.2 & 8.5  & 12.1 & 9.2 &7.1 \\ 
$\eta$ mass window cut & 7.1 & 5.2 & 5.3 & 19.8 & 9.6 & 6.1 & 12.1  & 8.5 & 3.1 &3.6 \\ 
$\eta^{\prime}$ mass window cut & 13.7 & 11.3 & 9.3 & 7.1 & 11.8 & 0.3 & 19.8 & 9.1 & 8.5 &4.7 \\ 
Q-factor method  & 12.7 & 14.1 & 6.6 & 4.7 & 6.8 & 1.9 & 17.7 & 8.8& 1.4 & 7.2\\ 
Multi-dimensional reweighting method & 15.2 & 13.8 & 6.5 & 6.6 & 9.3 & 0.9 & 7.7 & 3.9& 4.8 & 11.2\\ 
Resonance parameters& 12.2& 6.5 & 4.7 & 6.5 & 0.9 & 7.1 & 15.0 & 0.5 & 15.2 & 11.3\\ 
Extra resonances& $^{+23.7}_{-3.4}$ & $^{+20.7}_{-17.1}$ & $^{+17.8}_{-28.5}$ & $^{+13.7}_{-15.7}$ & $^{+0.0}_{-24.7}$ & $^{+12.4}_{-11.7}$  & $^{+25.3}_{-17.0}$ & $^{+11.7}_{-26.8}$ & $^{+27.6}_{-4.9}$ &  $^{+25.9}_{-3.2}$ \\ \hline

Total & $^{+38.3}_{-30.3}$ & $^{+32.7}_{-30.6}$ & $^{+24.3}_{-33.0}$& $^{+30.6}_{-31.6}$ & $^{+22.7}_{-33.6}$ & $^{+17.2}_{-16.7}$ & $^{+43.1}_{-38.9}$ & $^{+23.3}_{-33.5}$ & $^{+34.6}_{-21.4}$ &$^{+32.8}_{-20.3}$ \\
\hline\hline
\end{tabular}
\end{center}
\end{table*}

\section{Summary}
To investigate the structures in the $\gamma\phi$ system of $\jpsi\ar\gamma\gamma\phi$~\cite{refkang}, a PWA of this decay is performed for the first time, by using a sample of $(10087\pm44)\times10^{6}$ $\jpsi$ events collected with the BESIII detector. The PWA result shows that the $\jpsi\ar\gamma R, R\ar\gamma\phi$ process has predominantly $2^{++}$ components. The decays of $\eta(1405)$, $X(1835)$, several $f$-states, and $\eta_{c}$ into $\gamma\phi$ are observed with statistical significance greater than 5$\sigma$. The $f$-states are associated with $f_{1}(1285)$, $f_{1}(1420)$, $f_{1}(1510)$, $f_{2}(1525)$, $f_{2}(1950)$, $f_{2}(2010)$, and $f_{0}(2200)$. The decays of $f_{1}(1285)$, $f_{1}(1420)$, $\eta(1405)$, and $X(1835)$ to the $\gamma\phi$ final state have been confirmed, while the decays of the other resonances to the $\gamma\phi$ final state have been observed for the first time. 

In the $\gamma\phi$ system, only one pseudoscalar particle, $\eta(1405)$, is identified with a mass around 1.4 GeV/$c^{2}$. The measured product BF of $\eta(1405)$ is $\mathcal{B}(\jpsi\ar\gamma\eta(1405)\ar\gamma\gamma\phi)=(3.57\pm0.18^{+0.59}_{-0.61})\times 10^{-6}$. Due to the lack of direct coupling between glueballs and photons, the glueball decay into $\gamma\phi$ is suppressed. The $\eta(1405)$ observed in $\jpsi\ar\gamma\gamma\phi$ is unfavorable to be a glueball, making it more likely to be an excited state of $\eta^{\prime}$. Furthermore, the significance of an additional $\eta(1475)$ is evaluated to be 3.7$\sigma$, indicating that the presence of the $\eta(1475)$ cannot be excluded.

The $J^{PC}$ of $X(1835)$ is  confirmed to be $0^{-+}$. The mass, width and product of BF of $X(1835)$ are determined to be $(1849.3 \pm 3.0 ^{+7.6}_{-10.0})$~MeV/$c^{2}$, $(179.6 \pm 8.7^{+22.5}_{-27.9})$~MeV and $ \mathcal{B}(\jpsi\ar\gamma X(1835)\ar\gamma\gamma\phi)=(3.37\pm0.19^{+0.78}_{-1.10})\times 10^{-6}$, respectively. The observation of the decay $X(1835)\ar \gamma\phi$  indicates that the $X(1835)$ is a second radial excited state of $\eta^{\prime}$ containing $s\bar{s}$ quarks~\cite{ref8,ref9}.

Observation of the $\eta_{c}\ar\gamma\phi$ decay, which is the first observation of the radiative decay of $\eta_{c}$, is important for studying the properties of charmonium, such as the decays of $\jpsi$ and $\chi_{cJ}$~\cite{etac,etac2}.

No evidence of $\eta(1295)$, $\eta_{1}(1855)$ or $X(2370)$ is observed in the process of $\jpsi\ar\gamma\gamma\phi$. The possible existence of $\eta(1295)$ cannot be excluded yet, because the production rate of $\eta(1295)$ in $\jpsi$ radiative decays is predicted to be suppressed due to SU(3) flavor symmetry~\cite{1295xx}. The measured upper limit on the product BF of $\jpsi\ar\gamma X(2370)\ar\gamma\gamma\phi$ and $\mathcal{B}(\jpsi\ar\gamma\eta_{1}(1855)\ar\gamma\gamma\phi)$ does not contradict the theoretical expectations for the proposed natures, pseudoscalar glueball for X(2370) and hybrid state for $\eta_{1}(1855)$, as discussed in the introduction.

The decays of $f_{1}(1510)$, $f_{2}(1525)$, $f_{2}(1950)$, $f_{2}(2010)$ and $f_{0}(2200)$ to $\gamma\phi$ are observed for the first time. Accurate measurement of the $R\ar\gamma\phi$ decay provides crucial theoretical input for understanding the properties of these particles~\cite{ref12,2370,Rad1}. 


\textbf{Acknowledgement}

The BESIII Collaboration thanks the staff of BEPCII and the IHEP computing center for their strong support. This work is supported in part by National Key R\&D Program of China under Contracts Nos. 2020YFA0406300, 2020YFA0406400; National Natural Science Foundation of China (NSFC) under Contracts Nos. 11635010, 11735014, 11835012, 11922511, 11935015, 11935016, 11935018, 11961141012, 12022510, 12025502, 12035009, 12035013, 12061131003, 12192260, 12192261, 12192262, 12192263, 12192264, 12192265, 12221005, 12225509, 12235017, 12361141819; the Chinese Academy of Sciences (CAS) Large-Scale Scientific Facility Program; the CAS Center for Excellence in Particle Physics (CCEPP); Joint Large-Scale Scientific Facility Funds of the NSFC and CAS under Contract No. U1832207; CAS Key Research Program of Frontier Sciences under Contracts Nos. QYZDJ-SSW-SLH003, QYZDJ-SSW-SLH040; CAS Project for Young Scientists in Basic Research YSBR-101; 100 Talents Program of CAS; The Institute of Nuclear and Particle Physics (INPAC) and Shanghai Key Laboratory for Particle Physics and Cosmology; ERC under Contract No. 758462; European Union's Horizon 2020 research and innovation programme under Marie Sklodowska-Curie grant agreement under Contract No. 894790; German Research Foundation DFG under Contracts Nos. 443159800, 455635585, Collaborative Research Center CRC 1044, FOR5327, GRK 2149; Istituto Nazionale di Fisica Nucleare, Italy; Ministry of Development of Turkey under Contract No. DPT2006K-120470; National Research Foundation of Korea under Contract No. NRF-2022R1A2C1092335; National Science and Technology fund of Mongolia; National Science Research and Innovation Fund (NSRF) via the Program Management Unit for Human Resources \& Institutional Development, Research and Innovation of Thailand under Contract No. B16F640076; Polish National Science Centre under Contract No. 2019/35/O/ST2/02907; The Swedish Research Council; U. S. Department of Energy under Contract No. DE-FG02-05ER41374.


\end{document}